\documentclass[]{tMPH2e}
\usepackage{epstopdf}
\usepackage{multirow}
\usepackage{hyperref} 

\hypersetup{%
   pdfpagemode=None, 
   pdfstartpage=1,
   pdfstartview=FitH,
   pdfmenubar=true,
   pdftoolbar=true,
   colorlinks = true,
   linkcolor=blue,
   citecolor=blue,
   bookmarksopen=false
 }

\newcommand{\Fkt}[1]{\,\mathsf {#1}}

\def\openone{\leavevmode\hbox{\small1\kern-3.3pt\normalsize1}}

\ifx\Tr\renewcommand{\Tr}{\Fkt{Tr}} 
\else\newcommand{\Tr}{\Fkt{Tr}}
\fi

\begin{document}
\doi{10.1080/0026897YYxxxxxxxx}
 \issn{}
\issnp{}
\jvol{00}
\jnum{00} \jyear{2013} 

\markboth{M. Tomza et al.}{Interatomic potentials and spectroscopy of Rb$_2$}

\title{Interatomic potentials, electric properties, and spectroscopy \\of the 
ground and excited states of the Rb$_2$ molecule:
    \textit{Ab initio} calculations and effect of a non-resonant field\footnote[1]{
Dedicated to Professor Bretislav Friedrich on the occasion of his 60th birthday}
  }

\author{Micha\l~Tomza$^{a,b}$, Wojciech Skomorowski$^a$, Monika Musia\l{}$^c$, Rosario
  Gonz\'alez-F\'erez$^d$, Christiane P. Koch$^b$, and Robert
  Moszynski$^a$ 
  \\\vspace*{6pt}
  $^a$\textit{Faculty of Chemistry, University of Warsaw, Pasteura 1, 
    02-093 Warsaw, Poland}, 
  $^b$\textit{Theoretische Physik, Universit\"at Kassel, 
    Heinrich-Plett-Str. 40, 34132 Kassel, Germany}
  $^c$\textit{Institute of Chemistry, University of Silesia, Szkolna 9,
    40-006 Katowice, Poland} 
  $^d$\textit{Instituto ‘Carlos I’ de F\'isica
    Te\'orica y Computacional and Departamento de F\'isica At\'omica,
    Molecular y Nuclear, Universidad de Granada, 18071 Granada, Spain}
}
\maketitle
\begin{abstract}
  We formulate the theory for a diatomic molecule in a spatially
  degenerate electronic state interacting  
  with a non-resonant laser field and investigate its rovibrational 
  structure in the presence of the field.
  We report on \textit{ab initio} calculations employing the double
  electron attachment intermediate Hamiltonian Fock space coupled
  cluster method restricted to single and double excitations 
  for all electronic states of the Rb$_2$ molecule up to $5s+5d$ dissociation
  limit of about 26.000$\,$cm$^{-1}$. In order to correctly predict the
  spectroscopic behavior of Rb$_2$, we 
  have also calculated the electric transition dipole moments,
  non-adiabatic coupling and spin-orbit coupling matrix elements, and
  static dipole polarizabilities, using the 
  multireference configuration interaction method.
  When a molecule is
  exposed to strong non-resonant light, its rovibrational levels get
  hybridized. We study the spectroscopic signatures of this effect for
  transitions between the X$^1\Sigma_g^+$ electronic ground state and
  the A$^1\Sigma_u^+$ and  b$^3\Pi_u$ excited state manifold. The
  latter  is characterized by strong   perturbations due to the
  spin-orbit interaction. We find that for 
  non-resonant field strengths of the order $10^9$W/cm$^2$, the spin-orbit
  interaction and coupling to the non-resonant field become
  comparable. The non-resonant field can then be used to control the
  singlet-triplet character of a rovibrational level. 
\end{abstract}

\bigskip
\bigskip

\section{Introduction}
\label{sec:intro}

Rubidium was one of the first species to be Bose-condensed~\cite{Cornell96},
and nowadays it can routinely be cooled and trapped. It has therefore
become the drosophila of ultracold physics. Its long-range interatomic
interactions have extensively been studied,  and this has allowed to
very accurately determine the scattering length and $C_6$
coefficient~\cite{RobertsPRL98,vanKempenPRL02,MartePRL02}.
Rb$_2$ molecules have been formed out of ultracold rubidium atoms
using both photo- and
magneto-association~\cite{GabbaniniPRL00,DuerrPRL04}. Photoassociation
and Feshbach spectroscopy have also served to measure  the low-lying 
shape resonances of the rubidium
dimer~\cite{BoestenPRL96,BoestenPRA97,VolzPRA05}. 
Trapping rubidium in an optical lattice has facilitated studies of 
atom-molecule dark states~\cite{WinklerPRL05} and transferring the
molecules into their vibrational ground state~\cite{LangPRL08}. 
The Rb$_2$ molecule continues to draw attention in the context of the
coherent control of ultracold
collisions~\cite{WrightPRL05,WrightPRA06,WrightPRA07,PechkisPRA11,CariniPRA13}   
and femtosecond
photoassociation~\cite{SalzmannPRL08,MullinsPRA09,MerliPRA09,McCabePRA09}.  
These experiments as well as those employing photoassociation with
continuous wave
lasers~\cite{BergemanJPhysB06,HuangJPhysB06,HyewonPRA07,FiorettiJPB07,BellosPCCP11}   
require precise
spectroscopic knowledge not only of the  ground but also the
excited states for both interpretation and detection.  

The electronic ground and excited states have extensively been
studied. According to Huber and 
Herzberg~\cite{Huber:79}, the Rb$_2$ molecule was first observed in a
spectroscopic experiment by Lawrence and Edlefsen as early as
1929~\cite{LawrencePR29}. Cold molecules studies have lead to a
renewed interest in the Rb$_2$ molecule. The 
ground X$^1\Sigma_g^+$ state has been investigated in Ref.~\cite{LeRoy:00},
while the most  accurate experimental results for the a$^3\Sigma_u^+$
state have been reported by Lozeille et al.~\cite{Lozeille:06}, Beser et
al.~\cite{Beser:09} and Tiemann and
collaborators~\cite{Tiemann:10}. The most important excited states  
corresponding to the $^2$S+$^2$P dissociation limit, the A$^1\Sigma_u^+$ and 
b$^3\Pi_u$ states, have extensively been analysed in
Ref.~\cite{BergemanJPhysB06}.  
Less experimental information is available for other excited
states. Notably, the (1)$^3\Sigma_g^+$ state has been studied in 
Ref.~\cite{Mudrich:09}, and Ref.~\cite{Aubock:10} reports the
experimental observation of the (2)$^2\Pi_g$ state. The 
pure long-range state of $0_g^-$ symmetry, that is important for the 
photoassociation  of ultracold Rb atoms, has been analysed in
Ref.~\cite{GutterresPRA02}. Several of these experimental data were
successfully employed to derive empirical potentials that reproduce
the spectroscopic data with  the experimental accuracy, cf.  
Refs.~\cite{LeRoy:00,Tiemann:10} for the ground state X$^1\Sigma_g^+$
and Refs.~\cite{Beser:09,Tiemann:10} for the a$^3\Sigma_u^+$ potential.
The coupled manifold of the A$^1\Sigma_u^+$ and 
b$^3\Pi_u$ states was deperturbed by Bergeman and
collaborators~\cite{BergemanJPhysB06,SalamiPRA09} with the  
corresponding potential energy curves and spin-orbit coupling matrix
elements reported in Ref.~\cite{SalamiPRA09}. Potential energy curves
for other electronic states fitted to the experimental data are
older, cf. Ref.~\cite{Amiot:86} for  
the empirical potential energy curve of the (1)$^1\Pi_g$ state, and 
Refs.~\cite{Amiot:87} and~\cite{Amiot:90} for those of the
(2)$^1\Sigma_g^+$ state and (2)$^1\Pi_u$ states, respectively.

Given this extensive amount of experimental data, it is not surprising
that many theoretical calculations have tackled the ground and
excited states 
of the rubidium dimer. The first {\em ab initio} calculation on the Rb$_2$
molecule dates back to 1980 and was reported by Konowalow and 
Rosenkrantz~\cite{KonowalowACS82}. 
Three recent studies have 
reported \textit{ab initio} data of varying accuracy for the
potential energy  curves and in some cases further properties such as
couplings and transition moments of Rb$_2$. The 
non-relativistic potentials for all
molecular states by Park et al.~\cite{Park:01} show a 
root mean square deviation (RMSD) between the theoretical well depths
and the available experimental data of 235$\,$cm$^{-1}$, i.e., 9.9\%
on the average. The 2003 calculations by Edvardsson et al.~\cite{Edwardsson:03}
were devoted to the ground state potential and six excited state potentials of
ungerade symmetry. The spin-orbit coupling matrix elements were also
reported. The overall accuracy of these results was about the same as 
in Ref.~\cite{Park:01} with a RMSD of 180$\,$cm$^{-1}$ representing an
average error of 25\%. Note that since the number of states
considered in Refs.~\cite{Park:01} and~\cite{Edwardsson:03} differs,
the absolute RMSD may be smaller and the percentage error larger. Finally,
in 2012 Allouche and Aubert-Fr\'econ~\cite{Allouche:12} reported calculations
of all molecular states and spin-orbit coupling matrix elements corresponding
to the dissociation limits $5s+5s$, $5s+5p$, and $5s+4d$. These calculations
are much more accurate than any other previously reported in the literature with
a RMSD of 129 cm$^{-1}$, i.e., an error of 5.5\% only. However, they do not
cover highly excited molecular states that are of interest for conventional
spectroscopy experiments \cite{Jastrzebski}, for the
detection of ultracold molecules~\cite{Gould}
as well as photoassociation into
states with ion-pair character~\cite{BanCPL01,BanEPL04,TomzaPRA12}.

Photoassociation into highly excited electronic states is at the core
of a recent proposal for the production of ultracold Rb$_2$
molecules~\cite{TomzaPRA12}, aimed at improving earlier femtosecond
experiments~\cite{SalzmannPRL08,MullinsPRA09,MerliPRA09,McCabePRA09}. It
is  based on multi-photon transitions that can easily be driven by 
femtosecond laser pulses and allow to fully take advantage of the
broad bandwidth of femtosecond laser pulses while driving the narrow
photoassociation transition~\cite{KochFaraday09}. Moreover,
multi-photon photoassociation populates highly excited electronic
states with ion-pair character and strong spin-orbit
interaction. These 
features are advantageous for an efficient stabilization of the
photoassociated molecules into deeply bound molecules in the
electronic ground state~\cite{TomzaPRA12}. 
The theoretical modeling of the proposed photoassociation scheme
required the knowledge of precise {\em ab initio} potential energy
curves including those for highly excited states, spin-orbit and
nonadiabatic coupling matrix elements,  electric transition dipole
moments and dynamical Stark shifts. These data were not available in
the literature for the highly excited states, and the non-adiabatic
couplings and dynamical Stark shifts have been missing even for the
lowest states. Moreover,  
the newly developed tools of electronic structure
theory based on the Fock space coupled cluster method 
\cite{MusialJCP12,MusialRMP07,MusialCR12} could possibly allow for
reaching a better accuracy of the potentials than reported in 
Refs.~\cite{Park:01,Edwardsson:03,Allouche:12}. 
Last but not least, calculations of the electric properties for
diatomic molecules in spatially degenerate electronic states are scarce. 
To the best of our knowledge, only two studies considered this problem,
in the context of the dispersion interactions between
molecules~\cite{SpelsbergJCP99,SkomorowskiJCP11} rather than
non-resonant interactions with an external field, and a systematic
theoretical approach has not 
yet been proposed. Moreover, the presence of spin-orbit coupling between
the electronic states has been neglected in a recent treatment of
nuclear dynamics in a non-resonant
field~\cite{AganogluKoch,GonzalezPRA12}. Such an approximation does
not allow to study the competition between the spin-orbit coupling and 
the interaction with a non-resonant field which may both significantly
perturb the spectrum.

Here, we fill this gap and report the theoretical framework 
for a $\Pi$ state molecule interacting with a non-resonant
field and study its rovibrational dynamics in the presence of the
field. We also report {\em ab initio} calculations of all potential 
energy curves, spin-orbit and nonadiabatic coupling matrix elements
corresponding to the dissociation limits up to and including $5s+5d$. 
We test our {\em ab initio} results by comparing the main spectroscopic 
characteristics of the potentials to the available experimental data. 
We devote special emphasis to the
important manifold of the  A$^1\Sigma_u^+$ and b$^3\Pi_u$ states,
comparing our results to Refs.~\cite{BergemanJPhysB06,SalamiPRA09}. 
Since the electric properties of spatially degenerate electronic
states were not extensively studied in the literature thus far, we 
report here, to the best of our knowledge, the first {\em ab initio}
calculation of the irreducible components of  
the polarizability tensor, including their dependence on the interatomic
distance $R$, for the  A$^1\Sigma_u^+$ and b$^3\Pi_u$ states.
Finally, we study the effect of a non-resonant 
field on the spectroscopy in the A$^1\Sigma_u^+$ and b$^3\Pi_u$
manifold. This is motivated by our recent proposal for enhancing
photoassociation by controlling shape resonances with non-resonant
light~\cite{AganogluKoch,GonzalezPRA12}. In order to significantly
modify the scattering continuum of the atom pairs to be 
photoassociated, rather large non-resonant intensities are
required. Since the bound rovibrational levels are much more 
affected by a strong non-resonant field than continuum states, it
is important to investigate how the corresponding spectroscopic
features change.

Our paper is organized as follows. In Sec.~\ref{sec:2} we formulate
the theory of the interaction of a homonuclear molecule with an
external non-resonant field. In Sec.~\ref{sec:nonres} we provide the 
theoretical description of the perturbation of spectra by a 
non-resonant field, using as an example the spin-orbit coupled
manifold of the  A$^1\Sigma_u^+$ and b$^3\Pi_u$ electronic states
of Rb$_2$. We briefly summarize the {\em ab initio} methods employed
in our calculations in Sec.~\ref{sec:abinitio} and discuss the results
of these calculations in Sec.~\ref{sec:num}. In particular, we compare 
our data with results available in the literature and discuss 
the ability of the {\em ab initio} results to reproduce the 
high-resolution spectroscopic data for the A$^1\Sigma_u^+$ and
b$^3\Pi_u$ manifold~\cite{BergemanJPhysB06,SalamiPRA09}. We then
describe the 
interaction with a non-resonant field and study its spectroscopy
signatures on the transitions between the electronic ground state and
the A$^1\Sigma_u^+$ and b$^3\Pi_u$ manifold. Finally,
Sec.~\ref{sec:concl} concludes our paper.

\section{Diatomic molecule in a non-resonant electric field}
\label{sec:2}
We consider the interaction of a diatomic molecule with an
electric field with the direction taken along the $Z$ axis of the
space-fixed coordinate system, $\vec{\mathcal{E}}=(0,0,\mathcal{E})$.
To the second order, the Hamiltonian for the interaction of the
molecule with the electric field in the space-fixed frame can be written as,
\begin{equation}\label{eq:stark}
H_{\rm int} = -d_Z^{\rm{SF}} \mathcal{E} -\frac{1}{2}\alpha_{ZZ}^{\rm{SF}} \mathcal{E}^2\,,
\end{equation}
where $d_Z^{\rm{SF}}$ and $\alpha_{ZZ}^{\rm{SF}}$ denote the
appropriate components of the electric dipole
moment and electric dipole polarizability in the space-fixed frame.
Since we deal with a homonuclear molecule, only the second term of the above Hamiltionian
will be relevant in the present analysis. To evaluate the matrix elements of the
Hamiltonian in the electronic and rovibrational basis, we rewrite
$\alpha_{ZZ}^{\rm{SF}}$ in terms of the polarizability components in the
body-fixed frame. The $\alpha_{ZZ}^{\rm{SF}}$ dipole polarizability component
can be expressed in terms of space-fixed irreducible tensor
components $\alpha^{(l),{\rm SF}}_m$ \cite{HeijmenMP96},
\begin{equation}
  \alpha_{ZZ}^{\rm{SF}}=-\frac{1}{\sqrt{3}}\alpha^{(0),\rm{SF}}_0
  +\sqrt{\frac{2}{3}}\alpha^{(2),\rm{SF}}_0\,.
\end{equation}
For the irreducible tensor components, the transformation from the
space-fixed to the body-fixed coordinate system is given by
the rotation matrices $D^{(l)^\star}_{m,k}(\widehat{R})$,
\begin{equation}
\alpha^{(l),\rm{SF}}_m=\sum_{k=-l}^l D^{(l)^\star}_{m,k}(\widehat{R}) \;\; \alpha^{(l),\rm{BF}}_k \;.
\label{sfbf}
\end{equation}
Hence, we have
\begin{equation}
\label{sf2}
  \begin{split}
    \alpha^{(0),\rm{SF}}_0= D^{(0)^\star}_{0,0}(\widehat{R}) \;\; 
    \alpha^{(0),{\rm BF}}_0=\alpha^{(0),\rm{BF}}_0\;,\\
    \alpha^{(2),\rm{SF}}_0=
    \sum_{k=-2}^{2}D^{(2)^\star}_{0,k}(\widehat{R}) \;\; 
    \alpha^{(2),\rm{BF}}_k\;.
  \end{split}
\end{equation}
For simplicity, we omit the superscripts SF/BF in the rest of the paper
as from now we will use only the body-fixed quantities. We assume in this 
paper that the molecular axis defines the body-fixed $z$ axis.
For a diatomic molecule the set of the Euler angles $\widehat{R}$ can be chosen as
$\widehat{R}=(0,\theta,0)$, where $\theta$ is the angle between the molecular axis
and the space-fixed $Z$ axis. This particular choice of the Euler angles is consistent 
with the requirement that the space-fixed $Y$ and body-fixed $y$ axes coincide.
The other possible set would be $\widehat{R}=(3\pi/2,\theta,\pi/2)$ 
which correspond to the coincidence of the space-fixed $X$ and body-fixed $x$ axes.
Note that for our specific choice of the
Euler angles, the Wigner $D$ functions appearing in Eqs. (\ref{sf2})  reduce to:
\begin{equation}
D_{0,k}^{(l)^\star}(\phi,\theta,0)=\left[\frac{(l-k)!}{(l+k)!}\right]^{1/2}P_l^k(\cos\theta),
\end{equation}
where $P_l^k$  are the associated Legendre polynomials.
For any diatomic molecule, the non-zero irreducible components of the dipole polarizability
are $\alpha^{(0)}_0$ and $\alpha^{(2)}_0$.
In addition, for a diatomic molecule in a $\Pi$ electronic state, the
$\alpha^{(2)}_{-2}$ and $\alpha^{(2)}_{2}$ terms do not vanish.
They should be viewed as off-diagonal polarizability tensor components connecting
two degenerate electronic states, $|\Pi_{1}\rangle$ and $|\Pi_{-1}\rangle$, with opposite projection of
the total electronic orbital angular momentum on the molecular axis. See,
for instance, Eq. (16) of Ref.~\cite{SkomorowskiJCP11}.

The non-vanishing body-fixed polarizability components
are most conveniently expressed in terms of the Cartesian tensor elements $\alpha_{ii}$, $i=x,y,z$. Then
$\alpha^{(0)}_0$ is related to the trace of the polarizability,
\begin{equation}
  \label{eq:pol0}
  \alpha^{(0)}_0=-\frac{1}{\sqrt{3}}
  \left(\alpha_{xx}+\alpha_{yy}+\alpha_{zz}\right)\,,
\end{equation}
$\alpha^{(2)}_0$ to the anisotropy of the polarizability,
\begin{equation}
  \label{eq:pol2}
  \alpha^{(2)}_0=\frac{1}{\sqrt{6}}\left(2\alpha_{zz}-\alpha_{xx}-\alpha_{yy}\right)\,,
\end{equation}
and, for a molecule in a $\Pi$ electronic state, $\alpha^{(2)}_{-2}$ and
$\alpha^{(2)}_{2}$ reflect the difference between two perpendicular
components,
\begin{equation}
  \label{eq:pol22}
  \alpha^{(2)}_2= \alpha^{(2)}_{-2}=\alpha_{yy}-\alpha_{xx}.
\end{equation}

For a diatomic molecule in a $\Sigma$ state, the definitions of the Cartesian
components of the polarizability tensor $\alpha_{ii}$ in
Eqs.~(\ref{eq:pol0}) to (\ref{eq:pol22})
are unambigous. The $zz$ and $xx$ components are simply the parallel
and perpendicular components,
$\alpha_{\parallel}$ and $\alpha_{\perp}$, respectively. 
Thus, the irreducible tensor components appearing in
Eqs.~(\ref{eq:pol0}) to (\ref{eq:pol22}) are simply related to the trace
$\alpha$ and the anisotropy $\Delta\alpha$ of the polarizability tensor,
\begin{equation}
  \alpha^{(0)}_0=-\sqrt{3}\alpha, \; \; \; \; \; \; \; 
  \alpha^{(2)}_0=\frac{2}{\sqrt{6}}\Delta\alpha\,.
\label{relations}
\end{equation}
Obviously, for a $\Sigma$ state molecule the $xx$ and $yy$ components are equal,
and $\alpha^{(2)}_2=0$.

In the case of a molecule
in a degenerate electronic state ($\Pi$, $\Delta$ etc.) some caution is needed
when employing the Cartesian components  $\alpha_{ii}$,
since one has to specify the basis of the electronic states, in which
these quantities are expressed. Equation~\eqref{eq:pol22} assumes the
Cartesian components, $\alpha_{yy}$ and $\alpha_{xx}$, to be
calculated for the $|\Pi_x\rangle$ state.
However, the Cartesian basis $\{|\Pi_{x}\rangle,|\Pi_y\rangle\}$ for the $\Pi$ electronic state
is not convenient for the dynamical calculations, since the spin-orbit coupling
matrix elements are complex in this basis. Therefore, we prefer to use the spherical basis 
$\{|\Pi_{-1}\rangle,|\Pi_1\rangle\}$
for the $\Pi$ state over the Cartesian basis  $\{|\Pi_{x}\rangle,|\Pi_y\rangle\}$ since it
avoids complex quantities in the calculations and allows
for a simple adaptation of the Hund's case $(a)$ wave function to a given
symmetry of the rovibrational level. Therefore, we will use
the irreducible  polarizability components $\alpha^{(l)}_m$
rather than the Cartesian $\alpha_{ii}$.

Combining Eqs.~\eqref{eq:stark} to \eqref{eq:pol22} and making use of properties
of the rotation matrices $D^{(l)^\star}_{m,k}(\widehat{R})$, one
arrives at the following Hamiltonian for the interaction of the
homonuclear diatomic molecule with the static electric field,
\begin{equation}\label{eq:int}
  H_{\rm int} = 
  -\frac{\mathcal{E}^2}{2} 
  \left[-\frac{1}{\sqrt{3}} \alpha^{(0)}_0
    + \sqrt{\frac{2}{3}}\alpha^{(2)}_0 P_2^0(\cos\theta)
    + \frac{1}{6}\alpha^{(2)}_2 P_2^2(\cos\theta) + 
    4\alpha^{(2)}_{-2} P_2^{-2}(\cos\theta) 
  \right]\,.
\end{equation}

The above Hamiltonian is valid for any isolated electronic state of a
diatomic homonuclear molecule.
Albeit, the last two terms in this equation are relevant only
for molecules in a $\Pi$ electronic state.
Let us stress here that although this form of the Hamiltonian
seems a bit elaborate at first glance, it simplifies the
evaluation of the matrix elements in the symmetry-adapted basis set,
and it also avoids any ambiguities when employing the Cartesian polarizability
components for degenerate electronic states.
Equation~\eqref{eq:int}
also assumes the frequency of the non-resonant field to be far from
any resonance which allows for using the static polarizability and the
two-photon rotating-wave approximation. Such a field can be produced
for example by a carbon dioxide laser with a wavelength of about
10$\,\mu$m,

\section{Hamiltonian for the Rb$_2$ molecule in the manifold of
the coupled A$^1\Sigma_u^+$ and b$^3\Pi_u$ excited states
interacting with a  non-resonant field} 
\label{sec:nonres}
We construct the Hamiltonian for the nuclear motion in  Hund's case
$(a)$ coupling scheme with the primitive basis functions
$|n,\Lambda\rangle |S,\Sigma\rangle |J,\Omega,M \rangle$ that are
products of the electronic $|n,\Lambda\rangle$, electron spin
$|S,\Sigma\rangle$ and rotational $|J,\Omega,M \rangle$  functions.
Here, $j$ is the total angular momentum quantum number,
$S$ is the total electronic spin quantum number, $\Lambda$ and
$\Sigma$ are the projections of the total electronic orbital and total
electronic spin angular momenta onto the  molecular axis,  and $M$ is
the projection of the total  angular momentum onto the $Z$ space-fixed
axis.
$n$ labels the nonrelativistic dissociation limit of the molecular
state. We also define the projection of the total, electronic orbital
plus spin, angular momentum onto the molecular axis,
$\Omega=\Lambda+\Sigma$. For the coupled A$^1\Sigma_u^+$ and
b$^3\Pi_u$ manifold, we consider the rovibrational levels of the $e$
spectroscopic symmetry and odd parity. For simplicity,  any hyperfine
structure effects are neglected here. The properly symmetry-adapted
Hund's case $(a)$ wavefunctions read,
\begin{equation}\label{eq:states}
  \begin{split}
    |{\rm A}^1\Sigma_{0^+_u},J,M,e\rangle&=|{\rm A},0\rangle |0,0\rangle |J,0,M\rangle\;,\\
    |{\rm b}^3\Pi_{0^+_u},J, M, e\rangle&=\frac{1}{\sqrt{2}}\left[ 
      |{\rm b},1\rangle |1,-1\rangle |J,0,M \rangle-|{\rm b},-1\rangle |1,1\rangle |J,0,M \rangle\right]\;,\\
    |{\rm b}^3\Pi_{2_u}, J, M, e\rangle&=\frac{1}{\sqrt{2}}\left[  
      |{\rm b},1\rangle |1,1\rangle|J,2,M \rangle-|{\rm b},-1\rangle |1,-1\rangle |J,-2,M\rangle\right]\;.
  \end{split}
\end{equation}
The first two states have a projection of the total angular momentum
onto the molecular axis $|\Omega|=0$, while the third one has
$|\Omega|=2$. In the field-free case, the state with $|\Omega|=2$ is
decoupled from the states with $|\Omega|=0$, and it is not accessible
from the ground electronic state in the one-photon dipolar transitions considered
here. Consequently, the field-free model Hamiltonian $H_{0}$ describing the nuclear
motion in the manifol of the coupled  A$^1\Sigma_u^+$ and b$^3\Pi_u$ states can be
represented by following $2\times2$ matrix,
\begin{equation}
  \label{eq:0u+}
   H_{0} =
  \begin{pmatrix}
    T_R+\frac{\vec{j}^2}{2\mu R^2} + V^{{\rm A}^1\Sigma_u^+}(R) & \xi_1(R)\\
     \xi_1(R) & T_R+\frac{\vec{j}^2}{2\mu R^2}+V^{{\rm b}^3\Pi_u}(R)-\xi_2(R)
  \end{pmatrix} \,,
\end{equation}
where $T = T_R + \frac{\vec{j}^2}{2\mu R^2}$ is the sum of the vibrational
and rotational kinetic energy operators with $\vec{j}=\vec{J}-\vec{L}-\vec{S}$ being the mechanical angular momentum of the molecule and $V^k( R)$, $k={\rm A}^1\Sigma_u^+$, ${\rm b}^3\Pi_u$, denotes the
respective potential energy curves in the Born-Oppenheimer approximation. $\xi_1(R)=\langle
{\rm A}^1\Sigma_u^+|H_{\rm SO}|{\rm b}^3\Pi_u\rangle_{|\Omega|=0}$ and
$\xi_2(R)=\langle {\rm b}^3\Pi_u|H_{\rm SO}|{\rm b}^3\Pi_u\rangle_{|\Omega|=0}$
are the spin-orbit coupling matrix elements, and only the electronic states with $|\Omega|=0$
are included. Our model does not account for Coriolis-type angular couplings, i.e., the couplings
of the $\Omega=0$ states with $\Omega=1$ states because their effect on the rovibrational dynamics is negligible compared to
the spin-orbit couplings, the error of the electronic structure data and the influence of the weak non-resonant field.
It is not surprising due to large reduced
mass of Rb$_2$ molecules whose inverse enters all coupling matrix elements.

When the electric field is switched  on, the $\Lambda=1$ and $\Lambda=-1$ components of the
b$^3\Pi_u$ state are coupled. The coupling results form the off-diagonal polarizability tensor components 
in the Hamiltonian of Eq.~(\ref{eq:int}).
 Therefore, not only the interaction $H_{\rm int}$ from
Eq.~(\ref{eq:int}) has to be added to the Hamiltonian $H_{0}$ for the
A$^1\Sigma_u^+$ and b$^3\Pi_u$ states with $|\Omega|=0$, but also the
matrix (\ref{eq:0u+}) has to be extended so as to include the $|\Omega|=2$
component originating from the b$^3\Pi_u$ state since it has the $\Lambda$ projections exactly
opposite to those found in the state  with $|\Omega=0|$ while all other
quantum numbers are the same.
Hence, in the presence of the electric field the rovibrational levels
of the A$^1\Sigma_u^+$ and b$^3\Pi_u$ manifold are obtained by diagonalizing
the Hamiltonian represented by the following $3\times3$ matrix,
\begin{equation}
  \label{eq:Ham_in_field}
   H =
  \begin{pmatrix}
    T + W^{{\rm A}^1\Sigma_u^+}(R,\theta) & \xi_1(R) & 0\\
     \xi_1(R) & T+W^{{\rm b}^3\Pi_u}(R,\theta)-\xi_2(R) & W_{0/2}(R,\theta)\\
    0 & W_{0/2}(R,\theta) &T+W^{{\rm b}^3\Pi_u}(R,\theta)+\xi_2(R)
  \end{pmatrix} \,. 
\end{equation}
The diagonal elements of the interaction potentials incorporating  the
interaction with non-resonant field are given by,
\begin{equation}
  \label{eq:Wj}
  W^{k}(R,\theta) = V^{k}(R)+H_{\rm int}^k\,,
\end{equation}
where $k=$A${}^1\Sigma^+_u$ or b${}^3\Pi_u$ and $H_{\rm int}^k$ is given
by Eq. (\ref{eq:int}) for the electronic state labeled by $k$.
The  off-diagonal term due to the  non-resonant field,
$W_{0/2}(R,\theta)$, couples the $|\Omega|=0_u^+$ and
$|\Omega|=2_u$ components resulting from the b${}^3\Pi_u$ state. It
is proportional to the off-diagonal polarizability of the molecule in
the b$^3\Pi$ state,
\begin{equation}
  W_{0/2}(R,\theta)=-\frac{1}{12}\mathcal{E}^2 
  \alpha^{(2),{\rm b}^3\Pi_u}_2(R) P_2^2(\cos\theta)\,,
\end{equation}
with $\alpha^{(2)}_2$  defined by Eq.~(\ref{eq:pol22}). Analogously to
Eqs.~\eqref{eq:Ham_in_field} and \eqref{eq:Wj},
the Hamiltonian for the molecule in its electronic ground state
interacting with a non-resonant field
is simply given by $T+W^{{\rm X}^1\Sigma_g^+}(R,\theta)$.

\section{Ab initio electronic structure and dynamical calculations}
\label{sec:abinitio}

We adopt the computational scheme successfully applied
to the ground and excited states of the calcium dimer~\cite{BusseryPRA03,Moszynski:05,BusseryJCP06,BusseryMolPhys06,KochPRA08},
magnesium dimer~\cite{RybakPRL11,RybakFaraday11}, strontium dimer
\cite{SkomorowskiPRA12,SkomorowskiJCP12}, 
(BaRb)$^+$ molecular ion~\cite{KrychPRA11}, and SrYb
heteronuclear molecule~\cite{TomzaPCCP11}. 
The potential energy curves for 
the singlet  and triplet 
gerade and ungerade states
of the Rb$_2$ molecule corresponding to the  first seven lowest
dissociation limits, $5s+5s$, $5s+5p$, $5s+4d$, $5s+6s$, $5s+6p$,
$5p+5p$, and $5s+5d$, have been obtained by a supermolecule method,
\begin{equation}
V^{\rm ^{2S+1}|\Lambda|_{g/u}}(R)=
E_{\rm AB}^{\rm SM} -
E_{\rm A}^{\rm SM}-E_{\rm B}^{\rm SM}\,,
\label{cccv}
\end{equation}
where $E_{\rm AB}^{\rm SM}$ denotes
the energy of the dimer computed using the supermolecule method (SM),
and $E_{\rm X}^{\rm SM}$, 
$X=A$ or $B$, is the energy of the atom $X$ in the electronic state corresponding 
to the dissociation limit of the state ${\rm {}^{2S+1}|\Lambda|_{g/u}}$.
The full basis of the dimer was employed in the supermolecule calculations 
on the atoms $A$ and $B$, and the molecule $AB$, 
and the Boys and Bernardi scheme was utilized to
correct for the basis-set superposition error \cite{Boys:70}.
The calculations for the excited states employed the 
recently introduced Double Electron Attachment
Intermediate Hamiltonian Fock Space Coupled Cluster method restricted to
single and double excitations
(DEA-IH-FS-CCSD)~\cite{MusialJCP12,MusialRMP07,MusialCR12}. 
Starting with the closed-shell reference state for the doubly ionized
molecule Rb${}_2^{2+}$ that shows the  correct dissociation at large
interatomic separations, $R$, into closed-shell subsystems,
Rb$^+$+Rb$^+$, and using the double electron attachment 
operators in the Fock space coupled cluster ansatz makes our method
size-consistent at any interatomic separation $R$ and guarantee the
correct large-$R$ asymtptotics. Thus, the DEA-IH-FS-CCSD approach
overcomes the problem of the standard coupled cluster method
restricted to single and double excitations (CCSD) and the equation of
motion CCSD method~\cite{MusialRMP07} with the proper dissociation
into open-shell atoms. 
The potential energy curves obtained from the \textit{ab initio}
calculations were smoothly connected at intermediate interatomic separations with
the asymptotic multipole expansion~\cite{HeijmenMP96}. The $C_6$ coefficient of the
electronic ground state and the $C_3$ coefficient of the first excited
state were fixed at their empirical values derived from high-resolution
spectroscopic experiments~\cite{MartePRL02,GutterresPRA02}, while the
remaining coefficients were taken from Ref.~\cite{MarinescuPRA95}.

The transitions from the ground X$^1\Sigma_g^+$ state to the
$^1\Sigma_u^+$ and
$^1\Pi_u$ states and from the a$^3\Sigma_u^+$ to the $^3\Sigma^+_g$ and $^3\Pi_g$ states
 are electric dipole allowed.
The transition dipole moments for the electric
transitions were computed from the following expression~\cite{Bunker98},
\begin{equation}
d_i(n\leftarrow {\rm X})=\bigg\langle{\rm X}^1\Sigma_g^+\bigg|\hat d_i\bigg|
(n)^1|\Lambda|_u\bigg\rangle \; \; \; \; \; 
d_i(n\leftarrow {\rm a})=\bigg\langle{\rm a}^3\Sigma_u^+\bigg|\hat d_i\bigg|
(n)^3|\Lambda|_g\bigg\rangle, 
\label{trandipel}
\end{equation}
where the $\hat d_i$,  $i=x,y$ or $z$, denotes the $i$th component of the electric 
dipole moment operator.
Note that in the first term of Eq.~(\ref{trandipel}) $i=x$ or $y$ corresponds to transitions
to $^1\Pi_u$ states, while $i=z$ corresponds to transitions to $^1\Sigma_u^+$
states. The transitions from the a$^3\Sigma_u^+$ state
connect this state with the $^3\Pi_g$ and $^3\Sigma_g^+$ states, through the
$x$ and $y$ and $z$ operators, respectively.

We expect the rovibrational energy levels of the excited
electronic states of Rb$_2$ to show perturbations due to the nonadiabatic
couplings between the states. Analysing the pattern of the potential
energy curves,  we have found that many potential energy curves display
avoided crossings, suggesting strong radial couplings between these
electronic states. We have therefore computed the most
important radial coupling matrix elements, defined by the expression,
\begin{equation}
R({n\leftrightarrow n'})=
\left\langle(n)^{2S+1}|\Lambda|_{g/u}\left|\frac{\partial}{\partial R}
\right|(n')^{2S+1}|\Lambda|_{g/u}\right\rangle\,,
\label{radial}
\end{equation}
where $n\leftrightarrow n'$ signifies 
that the electronic states $n$ and $n'$ are coupled. 
Note that the radial derivative
operator couples states with the same projection of the
electronic orbital angular momentum on the molecular axis $\Lambda$.

Electric transition dipole moments, radial non-adiabatic coupling and
spin-orbit coupling matrix elements were obtained using the
Multireference Configuration Interaction method (MRCI) restricted to
single and double excitations with a large active space.
Scalar relativistic effects were included by using the small-core fully
relativistic energy-consistent pseudopotential ECP28MDF~\cite{LimJCP05}
from the Stuttgart library. Thus, in the present study
the Rb$_2$ molecule was treated as a system of effectively 18
electrons. The $[14s14p7d6f]$ basis set was employed in all
calculations. This basis 
was obtained by decontracting and augmenting the basis set of
Ref.~\cite{LimJCP05} by a set of additional functions improving the
accuracy of the atomic excitation energies of the rubidium atom with
respect to the NIST database~\cite{NIST}. 
The DEA-IH-FS-CCSD calculations were done with the code based on
the \textsc{ACES II} program system~\cite{ACESII}, while the MRCI
calculations were performed with the \textsc{MOLPRO} code~\cite{Molpro}.
All {\em ab initio} results reported in the present paper are available
from the Authors on request.

The rovibrational levels of the A$^1\Sigma_u^+$ and b$^3\Pi_u$ excited
state manifold are computed by diagonalizing the Hamiltonian (\ref{eq:0u+})
represented on a mapped Fourier grid, employing about $N_R=512$
radial grid points. For the calculations in the field 
we complement our Fourier grid representation for the radial
part by a basis set expansion in terms of Legendre polynomials for the
angular part, taking advantage of the magnetic quantum number $m$
being conserved.  We find that $j_{max}=19$ is sufficient to obtain
converged results for  ${\cal I}\le 2\times 10^{9}\,$W/cm$^2$. 
Presence of an intense non-resonant field leads to strong
hybridization of the rovibrational levels, and an adiabatic
separation of rotational and vibrational motion is not
applicable~\cite{AganogluKoch,GonzalezPRA12}. We account for this fact
by diagonalizing the full two-dimensional Hamiltonian, Eq.~(\ref{eq:Ham_in_field}),
represented by a $3 N_R (j_{max}+1) \times 3 N_R (j_{max}+1)$ matrix. 
For ${\cal I} \neq 0$,  the non-resonant field mixes
different partial waves, and $j$ and $j'$ are not good quantum numbers
anymore. For the sake of simplicity, we label the field-dressed
rovibrational levels
by the field-free quantum numbers. Note that the field-dressed levels
are adiabatically connected to their field-free counterparts even for
very large intensities.

\section{Numerical results and discussion}
\label{sec:num}
\subsection{Potential energy curves}
\label{sec:curves}

\begin{table*}[tbp]
\caption{Asymptotic enrgies (in cm$^{-1}$) and molecular states
  arising from different states of rubiudium
  atoms~\cite{Huber:79}. \label{tab1}} 
\begin{tabular}{lrrl}
\toprule
asymptote    & energy & energy &  molecular
states \\ 
& (present) & (exp.) & \\
\colrule
$(1)^2S(5s)$+$(1)^2S(5s)$   &  0      &    0   & $^1\Sigma_g^+$, $^3\Sigma_u^+$  \\
$(1)^2S(5s)$+$(1)^2P(5p)$   &  12731  &  12737 &  $^1\Sigma_g^+$, $^1\Pi_g$, $^1\Sigma_u^+$, $^1\Pi_u$, $^3\Sigma_g^+$, $^3\Pi_g$, \\
& & & $^3\Sigma_u^+$, $^3\Pi_u$ \\
$(1)^2S(5s)$+$(1)^2D(4d)$   &  19471  &   19355 & $^1\Sigma_g^+$, $^1\Sigma_u^+$, $^1\Pi_g$, $^1\Pi_u$, $^1\Delta_g$, $^1\Delta_u$,    \\
                            &       &         & $^3\Sigma_g^+$, $^3\Sigma_u^+$, $^3\Pi_g$, $^3\Pi_u$, $^3\Delta_g$, $^3\Delta_u$ \\
$(1)^2S(5s)$+$(2)^2S(6s)$   & 20126  & 20133 &  $^1\Sigma_g^+$, $^1\Sigma_u^+$, $^3\Sigma_g^+$, $^3\Sigma_u^+$\\
$(1)^2S(5s)$+$(2)^2P(6p)$   & 23732  & 23767 &   $^1\Sigma_g^+$, $^1\Pi_g$, $^1\Sigma_u^+$, $^1\Pi_u$, $^3\Sigma_g^+$, $^3\Pi_g$,\\
& & & $^3\Sigma_u^+$, $^3\Pi_u$ \\
$(1)^2P(5p)$+$(1)^2P(5p)$   & 25462  & 25475 &  $^1\Sigma_g^+$(2), $^1\Sigma_u^-$, $^1\Pi_g$, $^1\Pi_u$,  $^1\Delta_g$,\\
                            &       &       & $^3\Sigma_u^+$(2), $^3\Sigma_g^-$, $^3\Pi_g$, $^3\Pi_u$,  $^3\Delta_u$    \\
$(1)^2S(5s)$+$(2)^2D(5d)$   & 25736 & 25707 & $^1\Sigma_g^+$, $^1\Sigma_u^+$, $^1\Pi_g$, $^1\Pi_u$, $^1\Delta_g$, $^1\Delta_u$,    \\
                            &       &       & $^3\Sigma_g^+$, $^3\Sigma_u^+$, $^3\Pi_g$, $^3\Pi_u$, $^3\Delta_g$, $^3\Delta_u$ \\
Rb$^+(^1S)$+Rb$^-(^1S)$     & 29741 & 29771 & $^1\Sigma_g^+$, $^1\Sigma_u^+$\\
\botrule
\end{tabular}
\end{table*}
To test the ability of the {\em ab initio} approach adopted in the present
work to reproduce the experimental data, we first check the accuracy of
the atomic results. In Table~\ref{tab1} we report the excitation energies
at the dissociation limit computed with the DEA-IH-FS-CCSD method and compare 
the results to non-relativistic excitation energies
obtained with the Land\'e rule from the experimental excitation energies.
Inspection of Table~\ref{tab1}
shows that the agreement between the theoretical and experimental
excitation energies is very good. For the $5s+ns$ and $5s+np$ dissociation
limits,  the RMSD is only 21$\,$cm$^{-1}$, which represents an error of 0.08\%.
When the D states are included this good agreement is somewhat degraded.
The RMSD is now 49$\,$cm$^{-1}$, i.e., 0.26\%. This is due to the
lack of $g$ symmetry functions in the basis set used in our calculations.
Note parenthetically that we could not include $g$ functions in the basis,
because 
the ACESS II program does not support $g$ orbitals in the calculations
involving pseudopotentials. 
Our method reproduces very well the
electron affinity of the Rb atom, 3893$\,$cm$^{-1}$ on the theory side
vs. 3919$\,$cm$^{-1}$ measured in Ref.~\cite{FreyJPB78}, as well as the
ionization potential, 33630$\,$cm$^{-1}$ vs. 33690$\,$cm$^{-1}$~\cite{NIST}.
Finally, we note that the ground state static electric dipole
polarizability of the  atom obtained from our molecular calculations
is 319.5$\,$a$_0^3$ compared to 318.6$\,$a$_0^3$ from the most
sophisticated atomic calculations by Derevianko et
al.~\cite{Derevianko:10}. 

\begin{figure}[tbp]
  \begin{center}
    \includegraphics[width=0.75\columnwidth]{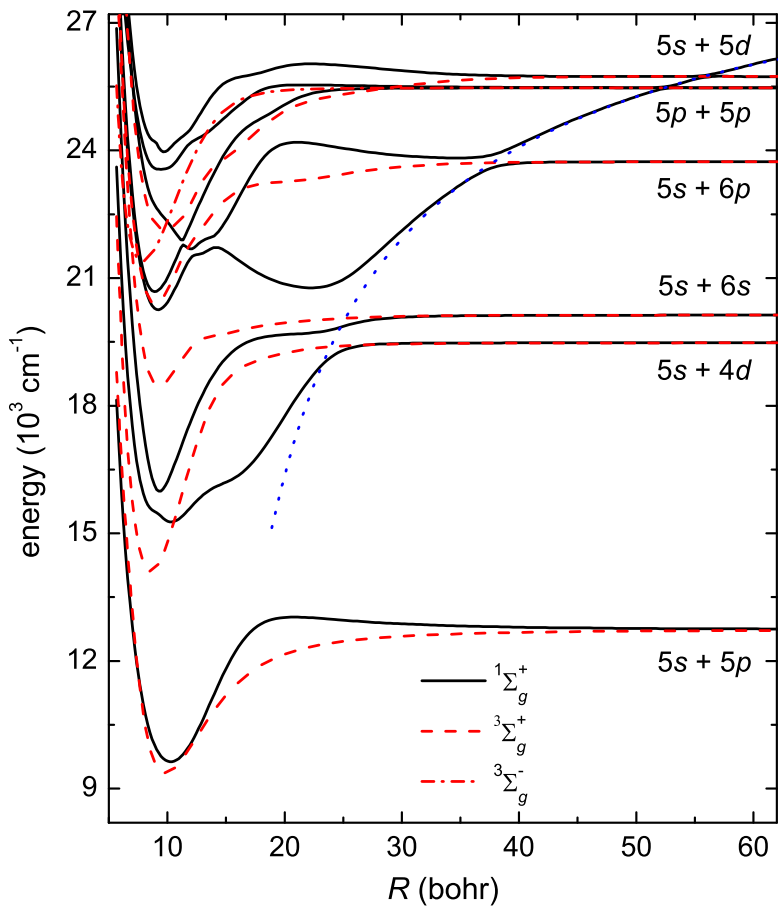}
  \end{center}
  \caption{Potential energy curves for the $^1\Sigma_g^+$ and
    $^3\Sigma_g^{\pm}$ states of  the Rb$_2$ molecule.} 
  \label{fig1}
\end{figure}
\begin{figure}[tbp]
  \begin{center}
    \includegraphics[width=0.75\columnwidth]{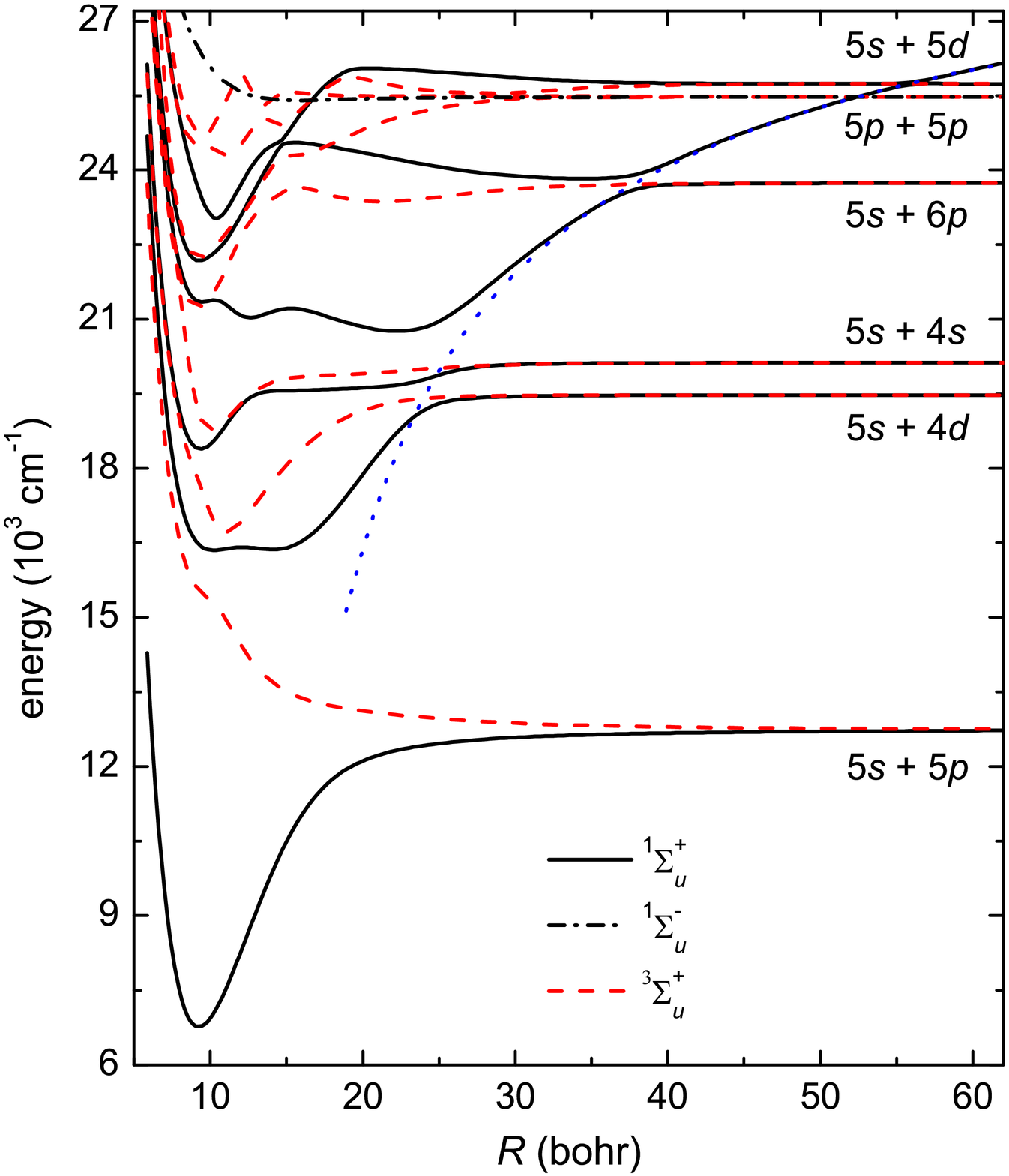}
  \end{center}
  \caption{Potential energy curves for the $^1\Sigma_u^{\pm}$ and
    $^3\Sigma_u^{+}$ states of  the Rb$_2$ molecule.} 
  \label{fig2}
\end{figure}
\begin{figure}[tbp]
  \begin{center}
    \includegraphics[width=0.75\columnwidth]{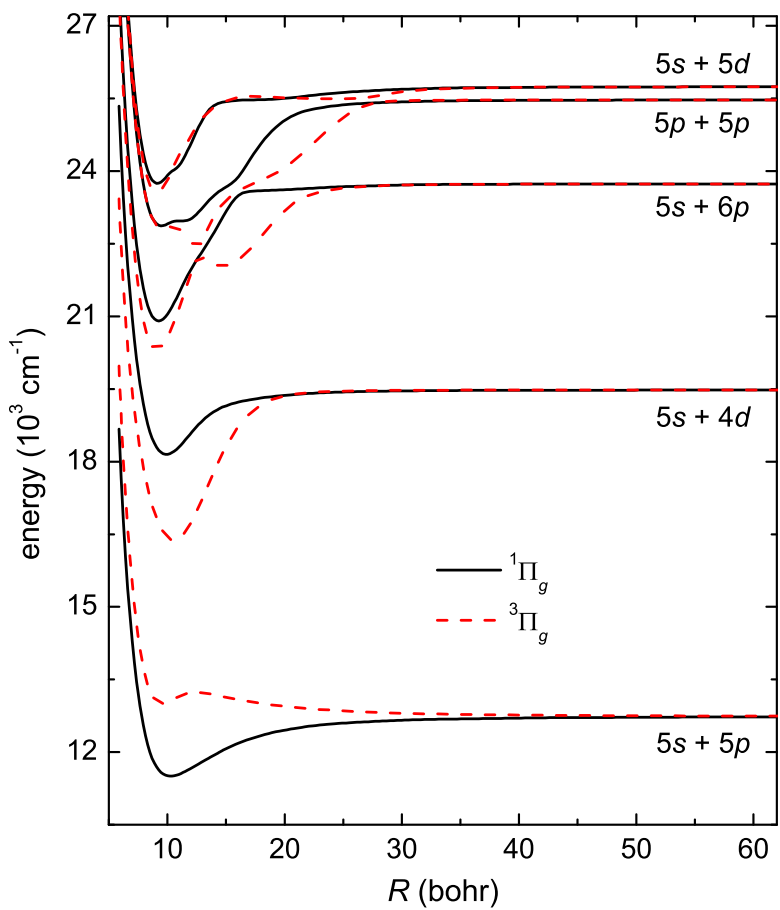}
  \end{center}
  \caption{Potential energy curves for the $^1\Pi_g$ and $^3\Pi_g$
    states of  the Rb$_2$ molecule.} 
  \label{fig3}
\end{figure}
\begin{figure}[tbp]
  \begin{center}
    \includegraphics[width=0.75\columnwidth]{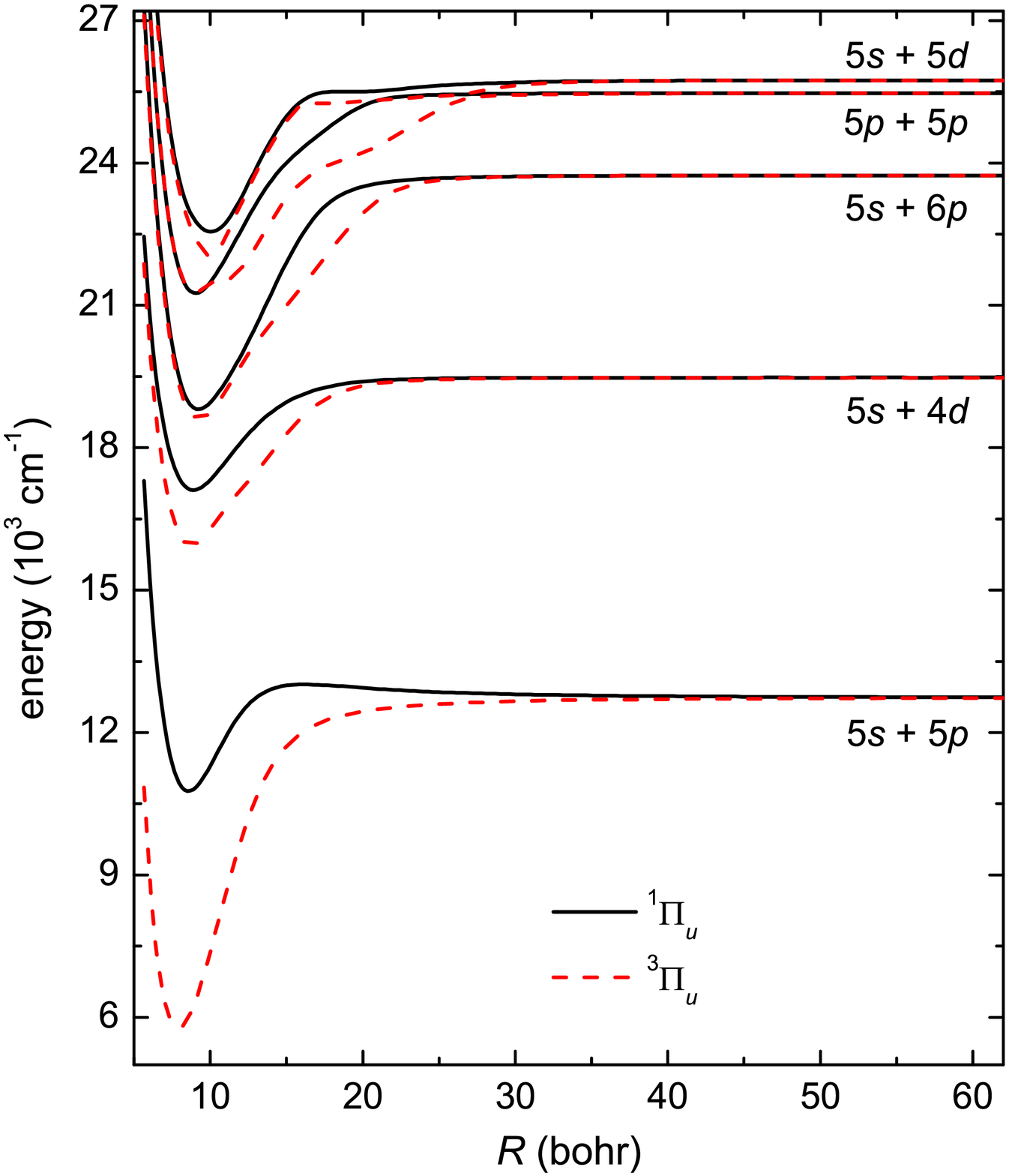}
  \end{center}
  \caption{Potential energy curves for the $^1\Pi_u$ and $^3\Pi_u$
   states of  the Rb$_2$ molecule.} 
  \label{fig4}
\end{figure}
\begin{figure}[tbp]
  \begin{center}
    \includegraphics[width=0.75\columnwidth]{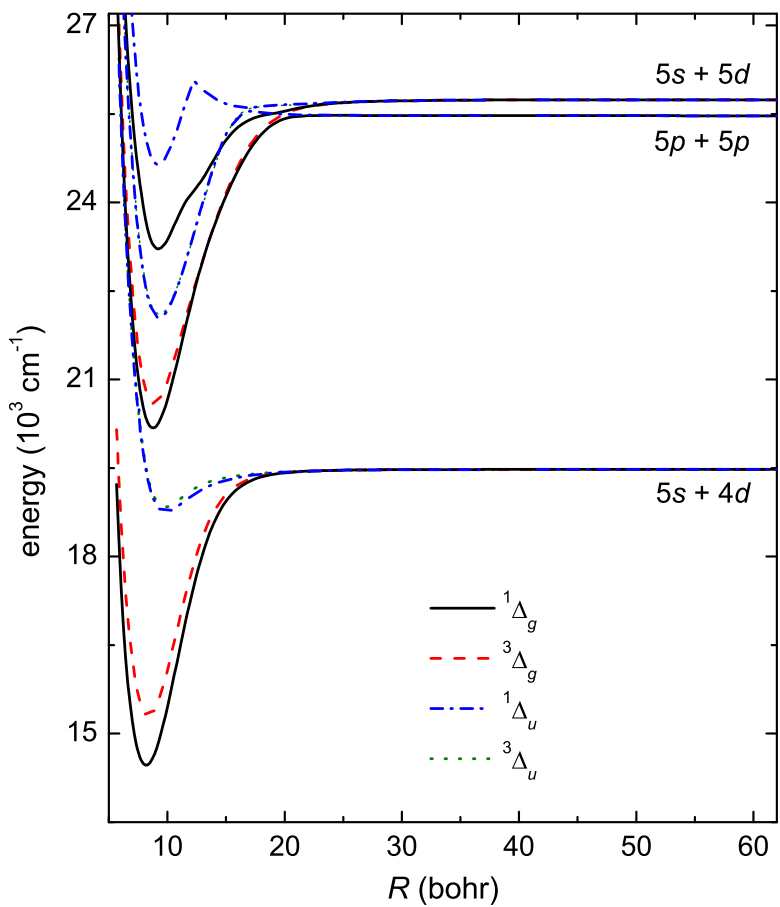}
  \end{center}
  \caption{Potential energy curves for the $^1\Delta_{g/u}$ and
    $^3\Delta_{g/u}$ states of  the Rb$_2$ molecule.} 
  \label{fig5}
\end{figure}
\begin{table*}[tbp]
  \caption{Spectroscopic characteristics of the non-relativistic
    $^1|\Lambda|_g$ electronic states  
    of ${}^{87}$Rb$_2$ molecule.\label{tab2}
  }
  \begin{tabular}{lcrrrrr}
\toprule
state           & Ref. & $R_e$ & $D_e$ &  $\omega_e$ & $T_e$ & asymptote \\
 & &  (bohr) & (cm$^{-1}$) &  (cm$^{-1}$) & (cm$^{-1}$) & \\
\colrule
X$^1\Sigma^+_g$   & present                 &  7.99 & 3912 & 56.1 &  0 &  $5s+5s$  \\
                  & \cite{LeRoy:00} (exp.)  &  7.96 & 3994 & 57.8  & 0 \\
                  & \cite{Allouche:12}      &  7.96 & 3905 & 58.4 & 0 \\
$(2)^1\Sigma_g^+$ & present                 & 10.29 & 3102 & 32.0 & 13545 & $5s+5p$\\
                  & \cite{Amiot:87} (exp.)  & 10.28 & 2963 & 31.5 & 13602 \\
                  & \cite{Allouche:12}      & 10.17 & 3084 & 31.2 & 13559 \\
$(3)^1\Sigma_g^+$ & present                 & 10.32 & 4210 & 32.9 & 19180 & $5s+4d$ \\
                  & \cite{Allouche:12}      & 10.20 & 4072 & 31.9 & 19189 \\
$(4)^1\Sigma_g^+$ & present                 &  9.34 & 4144 & 62.0 & 19898 & $5s+6s$ \\
$(5)^1\Sigma_g^+$ & present                 &  9.21 & 3483 & 37.8 & 24166 & $5s+6p$ \\
2nd. min.         & present                 & 22.22 & 2968 & 11.0 & 24681 &  \\
$(6)^1\Sigma_g^+$ & present                 &  8.93 & 3055 & 46.6 & 24594 & $5s+5p$ \\
2nd. min.         & present                 & 12.02 & 2056 & 50.6 & 25593 \\
3rd. min          & present                 & 34.60 &   86 &  4.7 & 27734 \\
$(7)^1\Sigma_g^+$ & present                 & 11.26 & 1852 & 92.8 & 25797 & $5s+5p$ \\
$(8)^1\Sigma_g^+$ & present                 &  9.47 &  183 & 41.3 & 27465 & $5s+5d$ \\
\colrule
$(1)^1\Pi_g$      & present                 & 10.25 & 1230 & 21.7 & 15417 & $5s+5p$\\
                  & \cite{Amiot:86} (exp.)  & 10.24 & 1290 & 22.3 & 15510 \\
                  & \cite{Allouche:12}      & 10.24 & 1198 & 22.0 & 15545 & \\
$(2)^1\Pi_g$      & present                 &  9.92 & 1326 & 31.0 & 22063 & $5s+4d$ \\
                  & \cite{Allouche:12}      &  9.88 & 1238 & 22.0 & 22023 & \\
$(3)^1\Pi_g$      & present                 &  9.25 & 2833 & 43.1 & 22149 & $5s+6p$ \\
$(4)^1\Pi_g$      & present                 &  9.48 & 2598 & 37.1 & 22099 & $5p+5p$ \\
$(5)^1\Pi_g$      & present                 &  9.13 & 1994 & 42.9 & 22187 & $5s+5d$ \\
\colrule
$(1)^1\Delta_g$   & present                 &  8.18 & 5026 & 48.7 & 18449 & $5s+4d$ \\
                  & \cite{Allouche:12}      &  8.14 & 4871 & 50.5 & 18390  \\
$(2)^1\Delta_g$   & present                 &  8.76 & 5291 & 57.6 & 24165 & $5p+5p$ \\
$(3)^1\Delta_g$   & present                 &  9.22 & 2528 & 56.5 & 27212 & $5s+5d$  \\
\botrule
\end{tabular}
\end{table*}
\begin{table*}[tbp]
  \caption{Spectroscopic characteristics of the non-relativistic $^3|\Lambda|_g$ electronic states 
    of ${}^{87}$Rb$_2$ molecule.\label{tab3}
  }
\begin{tabular}{lcrrrrr}
\toprule
state           & Ref. & $R_e$ & $D_e$ &  $\omega_e$ & $T_e$ & asymptote \\
 & &  (bohr) & (cm$^{-1}$) &  (cm$^{-1}$) & (cm$^{-1}$) & \\
\colrule
$(1)^3\Sigma_g^+$ & present                 &  9.91 & 3367 & 37.8 & 13279  & $5s+5p$\\
                  & \cite{Allouche:12}      &  9.73 & 3345 & 36.6 & 13298 \\
$(2)^3\Sigma_g^+$ & present                 &  8.58 & 5372 & 51.1 & 18017 & $5s+4d$ \\
                  & \cite{Allouche:12}      &  8.47 & 5347 & 51.5 & 17914 \\
$(3)^3\Sigma_g^+$ & present                 &  9.31 & 1657 & 38.2 & 22384 & $5s+6s$ \\
$(4)^3\Sigma_g^+$ & present                 &  8.95 & 3335 & 46.7 & 24313 & $5s+6p$ \\
$(5)^3\Sigma_g^+$ & present                 &  9.72 & 3488 & 19.4 & 26065 & $5p+5p$ \\
$(6)^3\Sigma_g^+$ & present                 &  9.19 & 3292 & 43.8 & 26953 & $5s+5p$ \\
$(7)^3\Sigma_g^+$ & present                 &  9.12 & 3268 & 38.5 & 27832 & $5s+5d$ \\
\colrule
$(1)^3\Pi_g$      & present                 &  9.54 & -267 & 30.3 & 16914 & $5s+5p$ \\
                  & \cite{Allouche:12}      &  9.47 & -268 & 30.3 & 16911 \\
$(2)^3\Pi_g$      & present                 & 10.56 & 3104 & 34.2 & 20285 & $5s+4d$ \\
                  & \cite{Allouche:12}      & 10.53 & 2927 & 33.6 & 20334 \\ 
$(3)^3\Pi_g$      & present                 &  9.08 & 3416 & 45.4 & 24232 & $5s+6p$ \\
$(4)^3\Pi_g$      & present                 &  9.06 & 2646 & 27.4 & 26735 & $5p+5p$ \\
$(5)^3\Pi_g$      & present                 &  9.09 & 2170 & 45.8 & 27484 & $5s+5d$ \\
\colrule
$(1)^3\Delta_g$   & present                 &  8.36 & 4181 & 48.3 & 19284 &  $5s+4d$  \\
                  & \cite{Allouche:12}      &  8.31 & 4017 & 48.9 & 19244 \\
$(2)^3\Delta_g$   & present                 &  8.85 & 5152 & 46.2 & 24588 &  $5s+5d$  \\
\botrule
\end{tabular}
\end{table*}
\begin{table*}[tbp]
  \caption{Spectroscopic characteristics of the non-relativistic
    $^1|\Lambda|_u$ electronic states  
    of ${}^{87}$Rb$_2$ molecule.\label{tab4}
  }
\begin{tabular}{lcrrrrr}
\toprule
state           & Ref. & $R_e$  & $D_e$ &
$\omega_e$ & $T_e$ & asymptote \\
&  &  (bohr) &  (cm$^{-1}$) & (cm$^{-1}$) & (cm$^{-1}$) & \\
\colrule
${\rm A}^1\Sigma_u^+$   & present                   &  9.24 & 5967 & 44.1 & 10680 & $5s+5p$\\
                  & \cite{SalamiPRA09} (exp.) &  9.21 & 5981 & 44.6 & 10750 &\\
                  & \cite{Allouche:12}        &  9.20 & 5896 & 44.4 & 10747 &   \\
$(2)^1\Sigma_u^+$ & present                   & 10.21 & 3128 & 20.5 & 20261 & $5s+4d$ \\
                  & \cite{Allouche:12}        & 10.09 & 3003 & 22.1 & 20258 \\
2nd. min          & present                   & 14.11 & 3112 & 13.5 & 20277 \\
                  & \cite{Allouche:12}        & 13.81 & 2926 & 11.5 & 20335 \\
$(3)^1\Sigma_u^+$ & present                   &  9.37 & 1737 & 42.4 & 22305 & $5s+6s$ \\
$(4)^1\Sigma_u^+$ & present                   &  9.46 & 2390 & 31.3 & 25258 & $5s+6p$ \\
2nd. min.         & present                   & 12.64 & 2702 & 24.3 & 24946 &  \\
3rd. min          & present                   & 22.26 & 2973 & 10.7 & 24675 & \\
$(5)^1\Sigma_u^+$ & present                   & 9.28  & 3565 & 39.1 & 26088 & $5p+5p$ \\
2nd. min          & present                   & 34.69 & 1920 &  5.0 & 27733 \\
$(6)^1\Sigma_u^+$ & present                   & 10.38 & 3308 & 52.9 & 26937 & $5s+5d$ \\
\colrule
$(1)^1\Pi_u$      & present                   &  8.57 & 1971 & 46.9 & 14676 & $5s+5p$ \\
                  & \cite{Amiot:90} (exp.)    &  -    & 1907 & 47.5 & 14666 \\
                  & \cite{Allouche:12}        &  8.48 & 1989 & 47.9 & 14654 \\
$(2)^1\Pi_u$      & present                   &  8.92 & 2369 & 31.6 & 21021 & $5s+4d$ \\
                  & \cite{Amiot:90} (exp.)    &  -    & 2454 & 36.4 & 20895 \\
                  & \cite{Allouche:12}        &  8.77 & 2157 & 36.1 & 21104 \\
$(3)^1\Pi_u$      & present                   &  9.23 & 4927 & 40.4 & 22721 & $5s+6p$ \\
$(4)^1\Pi_u$      & present                   &  9.03 & 4216 & 43.1 & 25166 & $5p+5p$ \\
$(5)^1\Pi_u$      & present                   & 10.06 & 3189 & 31.4 & 26465 & $5s+5d$ \\
\colrule
$(1)^1\Delta_u$   & present                   &  9.80 &  639 & 28.0 & 22825 &  $5s+4d$  \\
                  & \cite{Allouche:12}        &  9.78 &  542 & 26.9 & 22718 \\
$(2)^1\Delta_u$   & present                   &  9.31 & 3638 & 48.1 & 25818 &  $5p+5p$  \\
$(3)^1\Delta_u$   & present                   &  9.40 & 2630 & 34.2 & 27110 &  $5s+5d$  \\
\botrule\end{tabular}
\end{table*}
\begin{table*}[tbp]
  \caption{Spectroscopic characteristics of the non-relativistic 
    $^3|\Lambda|_u$ electronic states 
    of ${}^{87}$Rb$_2$ molecule.\label{tab5}
    }
\begin{tabular}{lcrrrrr}
\toprule
state           & Ref. & $R_e$  & $D_e$  &
$\omega_e$ & $T_e$ & asymptote \\ 
& & (bohr) & (cm$^{-1}$) & (cm$^{-1}$) & (cm$^{-1}$) & \\
\colrule
a$^3\Sigma_u^+$   & present                    & 11.46 &  250 & 13.5 & 3662 & $5s+5s$\\
                  & \cite{Tiemann:10} (exp.)   & 11.51 &  242 & 13.5 & - \\
                  & \cite{Allouche:12}         & 11.45 &  237 & 13.3 & 3669 \\
$(2)^3\Sigma_u^+$ & present                    & repulsive & -&- &-          & $5s+5p$\\
$(3)^3\Sigma_u^+$ & present                    & 11.02 & 2761 & 40.0 & 20628 & $5s+4d$\\
                  & \cite{Allouche:12}         & 10.96 & 2646 & 40.6 & 20614 \\
$(4)^3\Sigma_u^+$ & present                    & 10.06 & 1340 & 43.0 & 22701 & $5s+6s$\\
$(5)^3\Sigma_u^+$ & present                    &  9.18 & 2493 & 44.7 & 25155 & $5s+6p$\\
$(6)^3\Sigma_u^+$ & present                    &  9.29 & 3235 & 40.9 & 26147 & $5p+5p$\\
$(7)^3\Sigma_u^+$ & present                    &  9.09 &  938 & 47.2 & 28444 & $5p+5p$\\
\colrule
${\rm b}^3\Pi_u$        & present                    &  7.91 & 6969 & 57.2 &  9677 & $5s+5p$\\
                  &~\cite{SalamiPRA09} (exp.)  &  7.81 & 7039 & 60.1 &  9601 & \\
                  & \cite{Allouche:12}         &  7.88 & 7015 & 59.7 &  9624 & \\
$(2)^3\Pi_u$      & present                    &  8.73 & 3527 & 43.5 & 19862 & $5s+4d$\\
                  & \cite{Allouche:12}         &  8.60 & 3497 & 43.3 & 19764 & \\
$(3)^3\Pi_u$      & present                    &  9.28 & 5117 & 40.0 & 22531 & $5s+6p$\\
$(4)^3\Pi_u$      & present                    &  8.99 & 4189 & 43.3 & 25193 & $5p+5p$\\
$(5)^3\Pi_u$      & present                    & 10.04 & 3711 & 56.5 & 25943 & $5s+5d$\\
\colrule
$(1)^3\Delta_u$   & present                    &  9.83 &  719 & 27.3 & 22746 & $5s+4d$\\
                  & \cite{Allouche:12}         &  9.86 &  619 & 25.8 & 22641 \\
$(2)^3\Delta_u$   & present                    &  9.30 & 3695 & 40.7 & 25761 & $5s+5d$\\
\botrule
\end{tabular}
\end{table*}
The computed potential energy curves are reported in Fig.~\ref{fig1} for
the $^1\Sigma_g^+$ and $^3\Sigma_g^+$ symmetries, in Fig.~\ref{fig2} for 
the $^1\Sigma_u^+$ and $^3\Sigma_u^+$ symmetries, in Figs.~\ref{fig3} and
\ref{fig4} for the $^1\Pi_g$ and $^3\Pi_g$, and $^1\Pi_u$ and $^3\Pi_u$
symmetries, respectively. Finally Fig.~\ref{fig5} shows the
potential energy curves for the
singlet and triplet gerade and ungerade states of $\Delta$ symmetry.
The spectroscopic characteristics of the singlet gerade states are reported
in Table~\ref{tab2} while Table~\ref{tab3} collects these properties
for the triplet 
gerade states. Tables~\ref{tab4} and~\ref{tab5} present the 
results for the singlet and triplet states of ungerade symmetry, 
respectively. Inspection of Figs.~\ref{fig1} to~\ref{fig5} reveals that
almost all potential energy curves show a smooth behavior with well
defined minima. Some higher states display perturbations, mostly in the
form of avoided crossings, due to the interaction with other electronic
states of the same symmetry that are located nearby. At high energies the
density of states becomes so high that the avoided crossings produce
some irregularities in the curves. This is especially true for the
singlet and triplet gerade and ungerade states of  $\Sigma^+$ symmetry.
The $\Pi$ states show less perturbations, except for the avoided
crossings between the curves corresponding to the $(3)^1\Pi_g$ and
$(4)^1\Pi_g$, and $(3)^3\Pi_g$ and $(4)^3\Pi_g$ states. Interestingly,
the $\Pi_u$ states and the $\Delta$ states do not show any
irregularity due to nonadiabatic interactions between the states.

The agreement of the present potentials with those derived from the
experimental data is very good. This is demonstrated in
Tables~\ref{tab2} to~\ref{tab5}, where we compare the potential
characteristics 
with the available experimental data and with the most recent
calculations~\cite{Allouche:12}. For all the experimentally observed
states, the RMSD of our calculation is only 75.9$\,$cm$^{-1}$, i.e.,
the error is 3.2\%  on average, better than the most recent
calculations by Allouche and Aubert-Fr\'econ~\cite{Allouche:12} with
a RMSD of 129$\,$cm$^{-1}$ corresponding to an average error of
5.5\%. It is gratifying to observe that we reproduce low lying and
highly excited electronic states equally well. This is in a sharp 
contrast to Ref.~\cite{Allouche:12} which reproduces the well depth of
the $(2)^1\Pi_u$ state only with an error of 12\% compared to 3.5\%
for our calculation. Such a good agreement between theory
and experiment for the highest observed excited electronic state
gives us confidence that our predictions for the photoassociative
production of ultracold Rb$_2$ molecules in even higher electronic
states~\cite{TomzaPRA12} are  accurate.
Tables~\ref{tab2} to~\ref{tab5} also report the fundamental vibrational
frequencies $\omega_e$ for all electronic states considered in the
present paper. Except for the ground state, the agreement between
theory and experiment is within a few tenths of a wavenumber. Similar
agreement was found in the calculations by Allouche and
Aubert-Fr\'econ~\cite{Allouche:12}. 

\subsection{Non-adiabatic coupling  
  and spin-orbit coupling matrix elements} 
\label{sec:couplings}

\begin{figure}[tbp]
  \begin{center}
    \includegraphics[width=0.75\columnwidth]{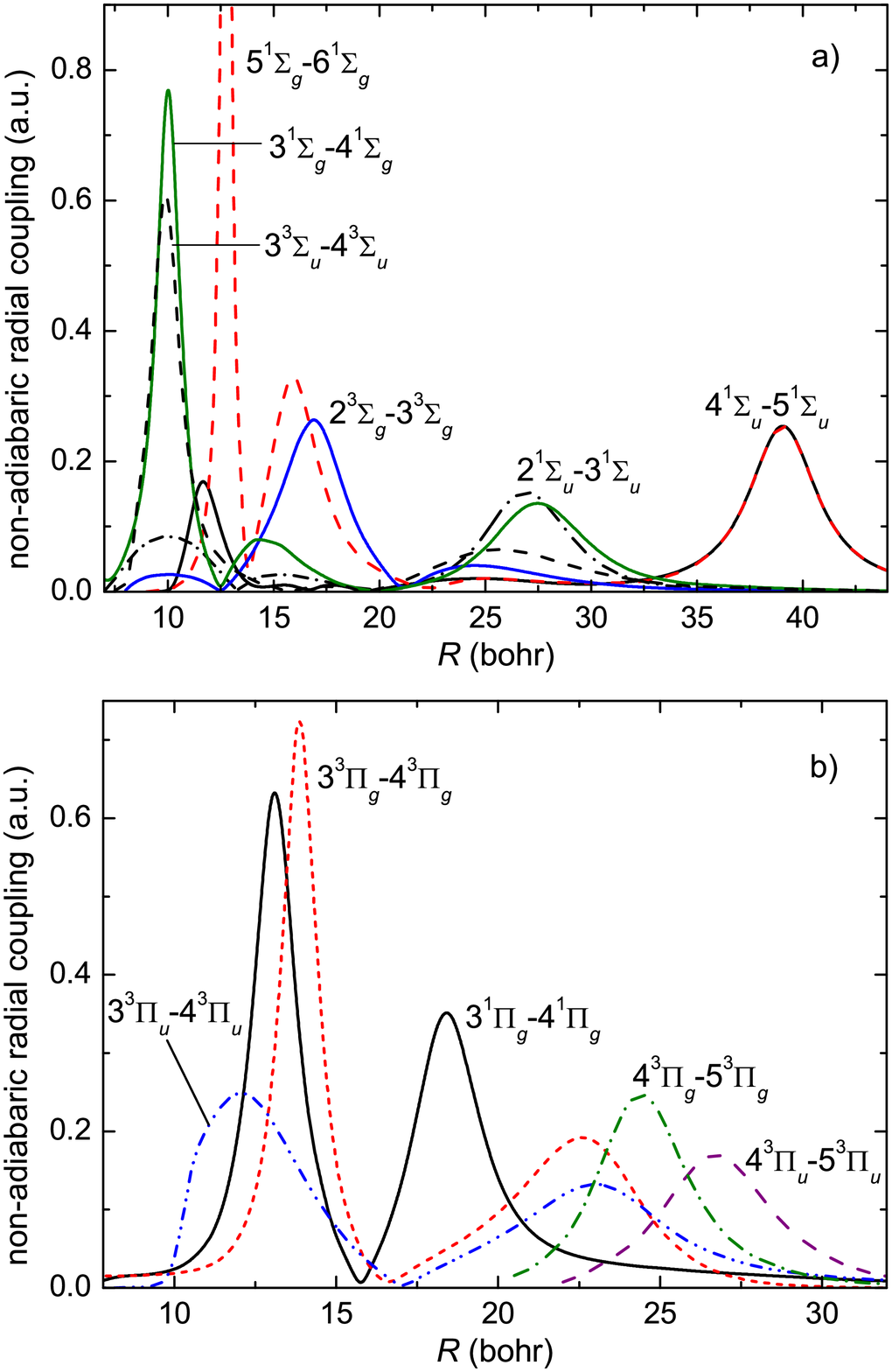}
  \end{center}
  \caption{Non-adiabatic radial coupling matrix elements between states
    of $\Sigma$ $(a)$ and $\Pi$ $(b)$ symmetry.} 
  \label{fig6}
\end{figure}
The importance of nonadiabatic interactions between electronic
states, resulting in the avoided crossings of the corresponding 
potential energy curves observed in Figs.~\ref{fig1} to~\ref{fig5},
can nicely be explained by analysing the
nonadiabatic coupling matrix elements computed according to
Eq.~(\ref{radial}). The nonadiabatic coupling matrix elements 
are reported in Fig.~\ref{fig6} for 
singlet and triplet states of $\Sigma_g^+$ and $\Sigma_u^+$ symmetry
(top) and the $\Pi$ states (bottom). As expected, the nonadiabatic coupling
matrix elements are smooth, Lorenzian-type functions, which, in the 
limit of an infinitely close avoided crossing, become a Dirac
$\delta$-function. The height and width of the curve depends on the
strength of the interaction. The smaller the width and the larger the
peak, the stronger is the interaction between the electronic states,
and the corresponding  potential energy curves are closer to each
other  at the avoided crossing. It is gratifying to observe that
the maxima on the nonadiabatic coupling matrix elements agree well
with the locations of the avoided crossing, and this despite the
fact that two very different methods were used in {\em ab initio}
calculations. Since the potential energy curves were shown to be
accurate, cf. the discussion in Sec. \ref{sec:curves},
we are confident that also the nonadiabatic coupling matrix elements
are essentially correct.

\begin{figure}[tbp]
  \begin{center}
    \includegraphics[width=0.75\columnwidth]{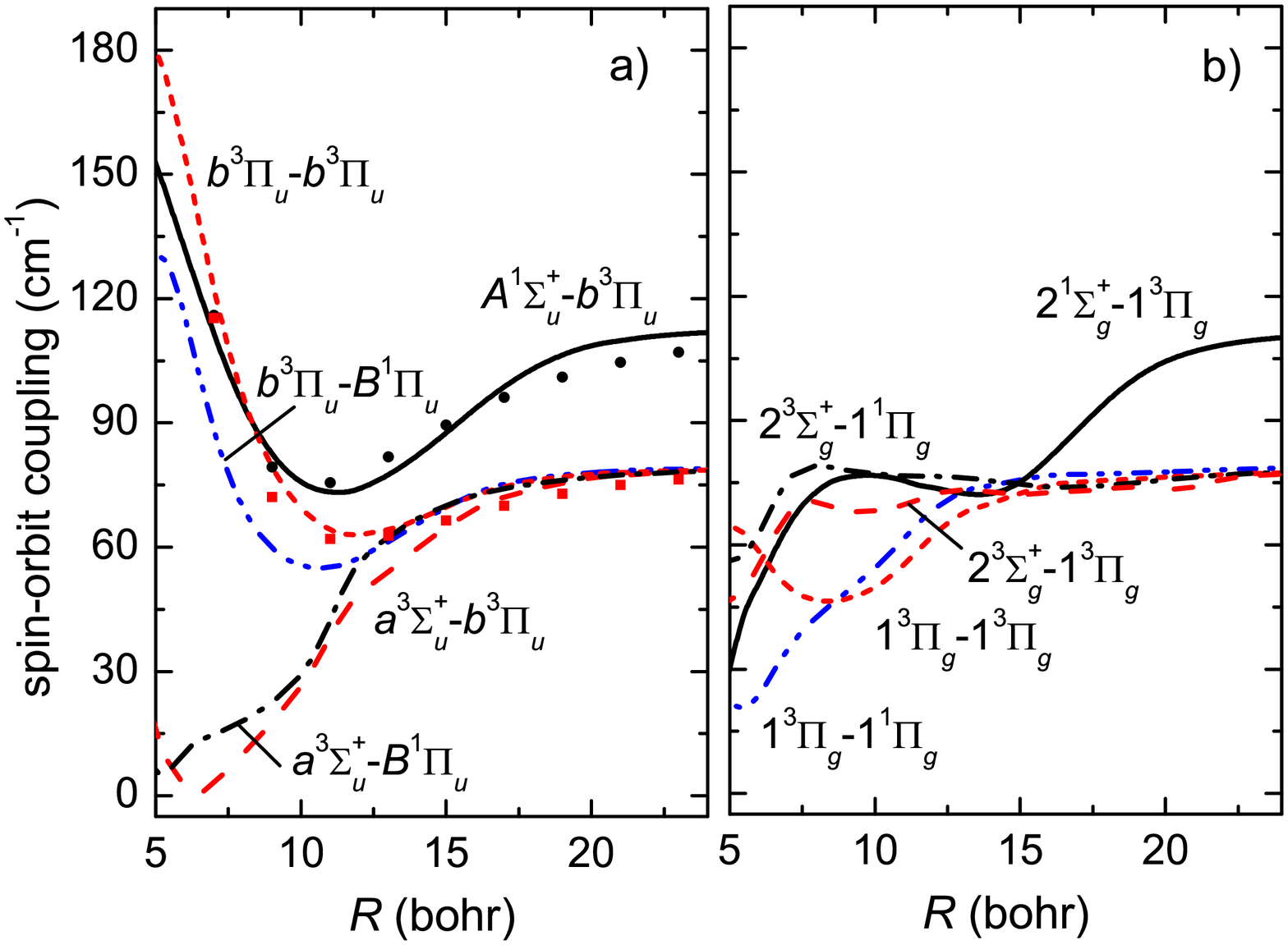}
  \end{center}
  \caption{Spin-orbit coupling matrix elements between states of
    ungerade $(a)$ and gerade $(b)$ symmetries dissociating into
    ${}^2S(5s)+{}^2P(5p)$. Black circles and red squares are analytical 
    fit to high-resolution spectroscopic data from Ref.~\cite{SalamiPRA09}.} 
  \label{fig7}
\end{figure}
Rubidium is a heavy atom and the electronic states of the Rb$_2$ molecule
show strong couplings due to the relativistic spin-orbit
interaction. Figure~\ref{fig7} reports the spin-orbit coupling matrix
elements as a function of the interatomic separation.  The matrix
elements are all 
represented by smooth curves approaching the atomic limit at large
$R$. The fine splittings of the atomic states are very accurately
reproduced by our calculations. For the first excited P state, 
the theoretical splitting between the 1/2 and 3/2 components is
236.2$\,$cm$^{-1}$ as compared to 237.6$\,$cm$^{-1}$ from the experiment.
It is also gratifying to observe that our {\em ab initio} calculations 
reproduce very well the spin-orbit coupling functions obtained from
fitting analytical functions to high-resolution spectroscopic data for
the A$^1\Sigma_u^+$ and b$^3\Pi_u$ manifold of
states~\cite{SalamiPRA09}. 
This gives us confidence that also perturbations in the molecular
spectra due to the spin-orbit interaction will  correctly be
reproduced from the present {\em ab initio} data.

\subsection{Electric transition dipole moments and electric dipole 
  polarizabilities} 
\label{subsec:transmom}

\begin{figure}[tbp]
  \begin{center}
    \includegraphics[width=0.75\columnwidth]{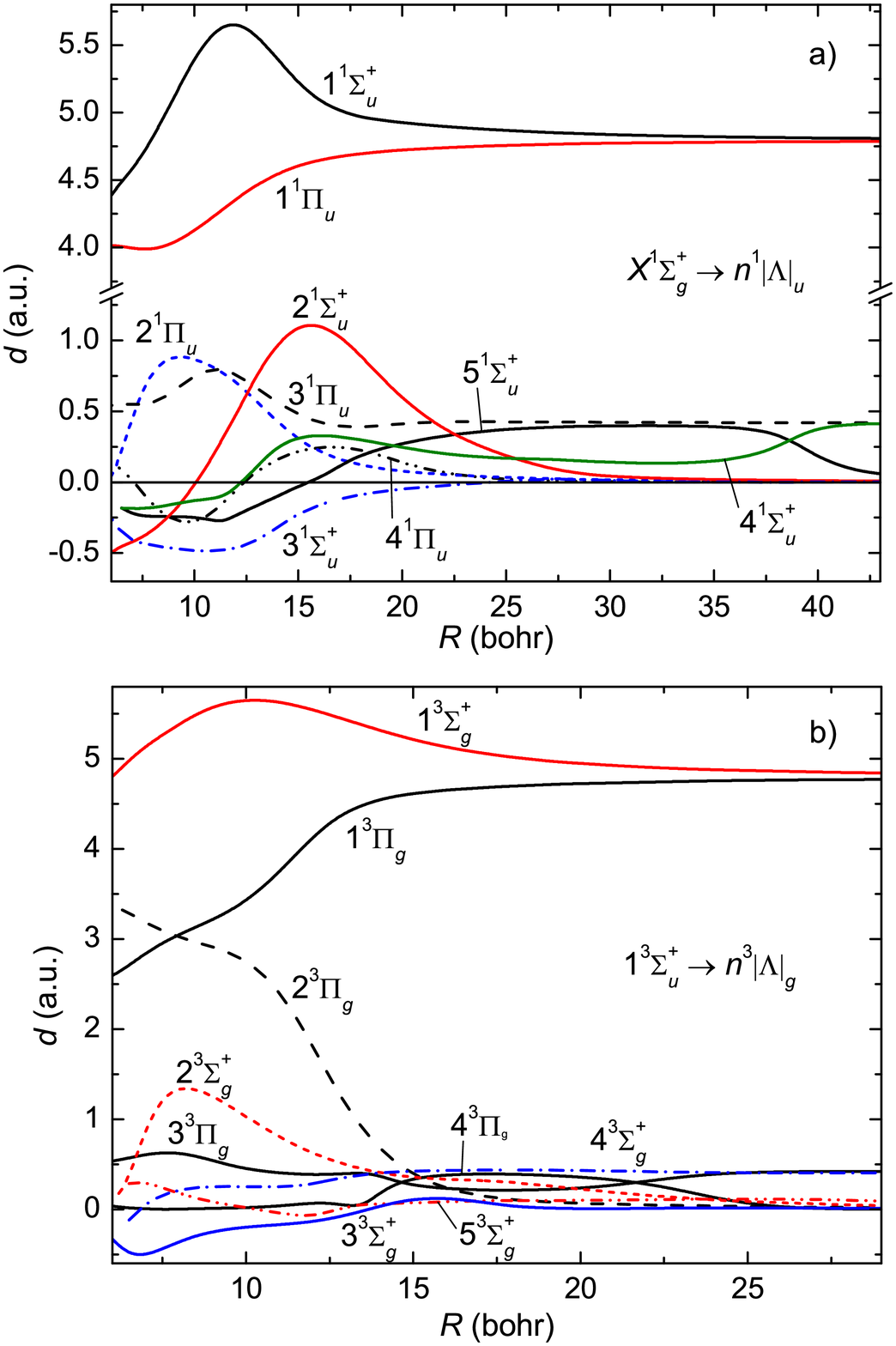}
  \end{center}
  \caption{Electric dipole transition moments: $(a)$ between the 
    X$^1\Sigma_g^+$ ground state and excited states of $^1\Sigma_u^+$
    and $^1\Pi_u$ symmetry and $(b)$ between the a$^3\Sigma_u^+$ lowest triplet 
    state and excited states of $^3\Sigma_g^+$ and $^3\Pi_g$
    symmetry.} 
  \label{fig8}
\end{figure}
\begin{figure}[tbp]
  \centering
  \includegraphics[width=0.75\columnwidth]{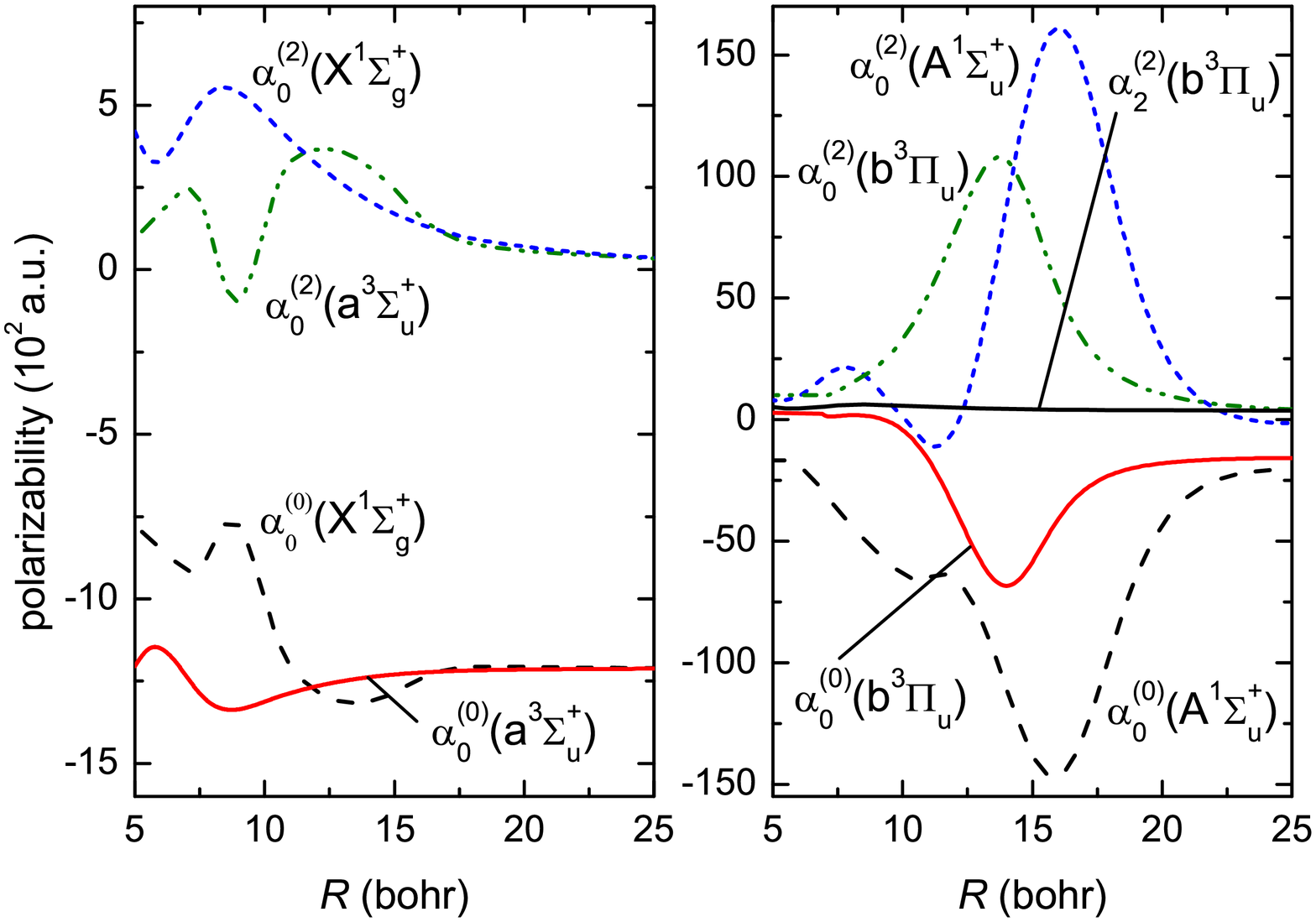}
  \caption{Electric dipole polarizabilities for the electronic ground
    state (left) and the first excited state (right).}
  \label{fig9}
\end{figure}

A full characterization of the molecular spectra requires knowledge 
of the electric transition dipole moments. These were calculated
according to Eq.~(\ref{trandipel}) and are presented in
Fig.~\ref{fig8} for transitions from the X$^1\Sigma_g^+$ ground  state 
and in Fig.~\ref{fig9} for transitions from the a$^3\Sigma_u^+$ lowest
triplet state. The strongest transitions
from the ground singlet state are those to the A$^1\Sigma_u^+$ and
$(1)^1\Pi_u$ states, i.e., to states corresponding to the first
excited dissociation limit. 
All other transition moments are much smaller, suggesting that the
corresponding line intensities in the spectra will be much weaker.
The same is true for transitions departing from the a$^3\Sigma_u^+$
state. The transition moments do not show a strong dependence on 
$R$, except at small interatomic separations, and smoothly
tend to their asymptotic atomic value.

The static electric dipole polarizabilities for the X$^1\Sigma_g^+$
electronic ground state, the a$^3\Sigma^+_u$ state, and the relevant
excited A$^1\Sigma_u^+$ and b$^3\Pi_u$ states are  
presented in Fig.~\ref{fig9}. They show an overall smooth behavior and
also tend smoothly to their asymptotic atomic values. The
interaction-induced variation of the polarizability is clearly visible
while changing the internuclear distance $R$. It is significant for
excited states, especially for the A$^1\Sigma_u^+$ state for which the
isotropic part $\alpha$ reaches 8000$\,a_0^3$, and the anisotropic
part $\Delta\alpha$ reaches 6000$\,a_0^3$. Such large values of both
the interaction-induced variation of isotropic and anisotropic
polarizabilities suggest  that the influence of the non-resonant laser
field on the rovibrational dynamics and transitions between the ground
X$^1\Sigma_g^+$ state, and the A$^1\Sigma_u^+$ and b$^3\Pi_u$ states  
should be significant even at relatively weak field intensities.
Comparing the present polarizabilities of the X$^1\Sigma_g^+$ and a$^3\Sigma_u^+$ 
states with theoretical results by Deiglmayr et al.~\cite{DeiglmayrJCP08},
we find good agreement. For example the isotropic polarizability $\alpha$ 
given by trace of the polarizability tensor for the  X$^1\Sigma_g$ and
a$^3\Sigma_u^+$ states being 522$\,$a.u. and 675$\,$a.u. in the present study 
and  533$\,$a.u. and 678$\,$a.u. in Ref.~\cite{DeiglmayrJCP08}, respectively.

Note parenthetically that the transition moments and matrix elements
of the spin-orbit coupling also change when a DC or non-resonant AC
field is applied, but the changes induced on the rovibrational
spectrum  are expected to be smaller compared to the effects
introduced within Eq.~\eqref{eq:Ham_in_field}. Therefore, the
investigation of the field-induced variation of the transition moments
and spin-orbit couplings is out of the scope of the present paper. 

\subsection{Rovibrational spectra in the ${\rm A}^1\Sigma_u^++{\rm b}^3\Pi_u$ manifold without a non-resonant field}
\label{subsec:nofield}

\begin{figure}[tbp]
  \begin{center}
    \includegraphics[width=0.75\columnwidth]{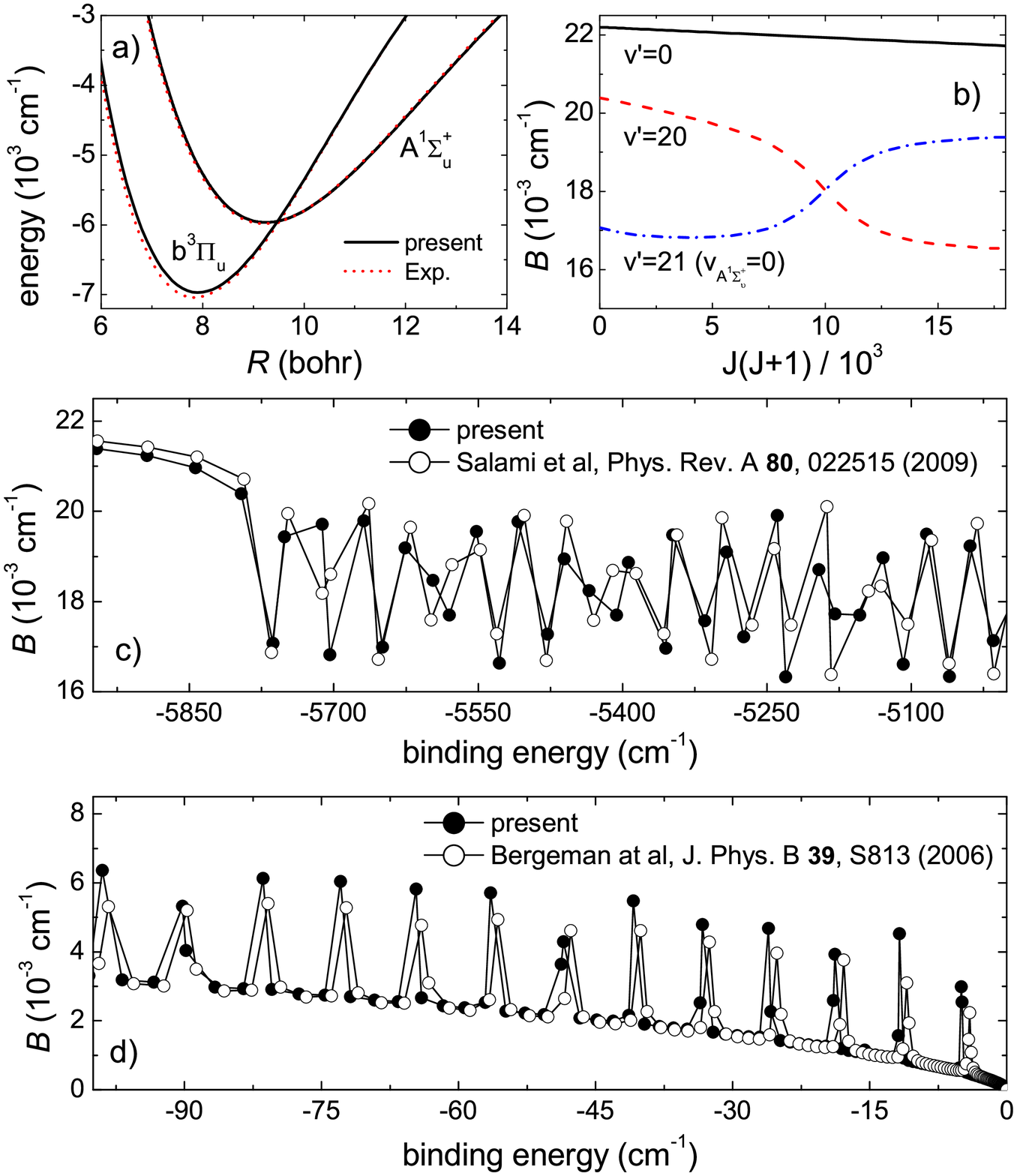}
  \end{center}
\caption{Characteristics of the rovibrational levels for the $\Omega=0_u^+$ component of the coupled A$^1\Sigma_u^+$ and b$^3\Pi_u$ manifold of states in $^{87}$Rb$_2$: $(a)$ present and empirical potential energy curves \cite{SalamiPRA09}, 
$(b)$ rotational spacings, and $j=1$ rotational constants for strongly bound levels $(c)$ and close to the
dissociation limit $(d)$.} 
\label{fig10}
\end{figure}

We now compare in more detail the ability of our {\em ab
  initio} data to reproduce the fine details of high-resolution
experiments of Ref.~\cite{SalamiPRA09}. In Fig.~\ref{fig10}$(a)$, 
we report the {\em ab initio} and empirical potentials for the
A$^1\Sigma_u^+$ and b$^3\Pi_u$ states of Rb$_2$. Inspection of 
Fig.~\ref{fig10}$(a)$ shows a very good agreement. The 
{\em ab initio} calculations reproduce the well depth of the
A$^1\Sigma_u^+$ state within 14 cm$^{-1}$ on the overall depth of 5981
cm$^{-1}$, i.e., within 0.2\%. 
The agreement for the b$^3\Pi_u$ state is slightly less good. The
difference in the well depths amounts to 70 cm$^{-1}$ for the well depth
of 7039 cm$^{-1}$. This represents an error of roughly 1\%. Such an
agreement between theory and experiment should be 
considered as very good. Also the crossing of the A$^1\Sigma_u^+$
and b$^3\Pi_u$ potential energy curves is perfectly reproduced. Our
dynamical calculations predict the level $v'=21$ to be the first
rovibrational level corresponding to the A state,  
see the rotational spacings in panel $(b)$ of Fig.~\ref{fig10}. This
is one quantum higher than predicted by the
experiment~\cite{SalamiPRA09}, but the 70 cm$^{-1}$  
disagreement in the well depths fully explains this difference.

Figure~\ref{fig10} also reports the rotational constants for the deeply bound 
rovibrational levels (panel $(c)$) and levels at the
threshold (panel $(d)$). Inspection of Fig.~\ref{fig10}$(c)$ 
reveals that theory correctly locates all levels that are not
perturbed by the spin-orbit interaction, and the first perturbed
level. The agreement in the rotational constants for the rovibrational
levels in the middle of the potential well is less good, but 
note the scale on the axis. Overall, we reproduce semi-quantitatively
the pattern of the rovibrational levels in this region of the
potentials. Also the oscillations of the rotational constants
reflecting the perturbations due to the spin-orbit coupling between
the  A$^1\Sigma_u^+$ and the b$^3\Pi_u$ states are correctly 
described. This is in accordance with the good agreement between the
{\em ab initio} spin-orbit coupling and the data fitted to the
experiment shown in Fig.~\ref{fig7}. 
The agreement of the rotational constants for the rovibrational levels
near the threshold is very good. This is partly due to the fact
that in our calculations we have used the best long-range coefficients
from atomic calculations~\cite{Derevianko:10}. 
However, the correct long-range coefficient alone would not be
sufficient to obtain such a good agreement between 
theory and experiment. In fact, panel $(d)$ of Fig. \ref{fig10} shows
that theory very precisely locates the repulsive walls of the
potentials near the zero crossing. This is very gratifying for a
theoretical calculation as this region of the potential 
energy curve is very difficult to describe with {\em ab initio} methods.

\subsection{Perturbation of the spectra by a non-resonant field} 
\label{subsec:nonres}
Bound rovibrational levels are strongly affected by a non-resonant
field~\cite{GonzalezPRA12}. We demonstrate in this section that not
only are the levels shifted in energy and is their rotational
motion strongly hybridized, but also, for levels in the coupled
A$^1\Sigma_u^+$ and b$^3\Pi_u$ excited state manifold, the
singlet-triplet composition may be changed. 
Note that the non-resonant field mixes different rotational and
possibly also vibrational states, and in the presence of the field, 
$v,j$, $v',j'$ are not good quantum numbers anymore. However, 
for simplicity, we do not distinguish between the field-free quantum
numbers $v,j$, $v',j'$ and the corresponding field-dressed labels
$\tilde v,\tilde j$, $\tilde v',\tilde j'$~\cite{GonzalezPRA12}. 
The carbon dioxide laser with wavelength of about 10$\,\mu$m is assumed
to be used as a source of a non-resonant field. For that wavelength, the static 
electric dipole polarizability is good approximation for the dynamic one with a few percent error  both 
for the ground and excited A+b states.

\begin{figure}[tbp]
  \centering
  \includegraphics[width=0.75\columnwidth]{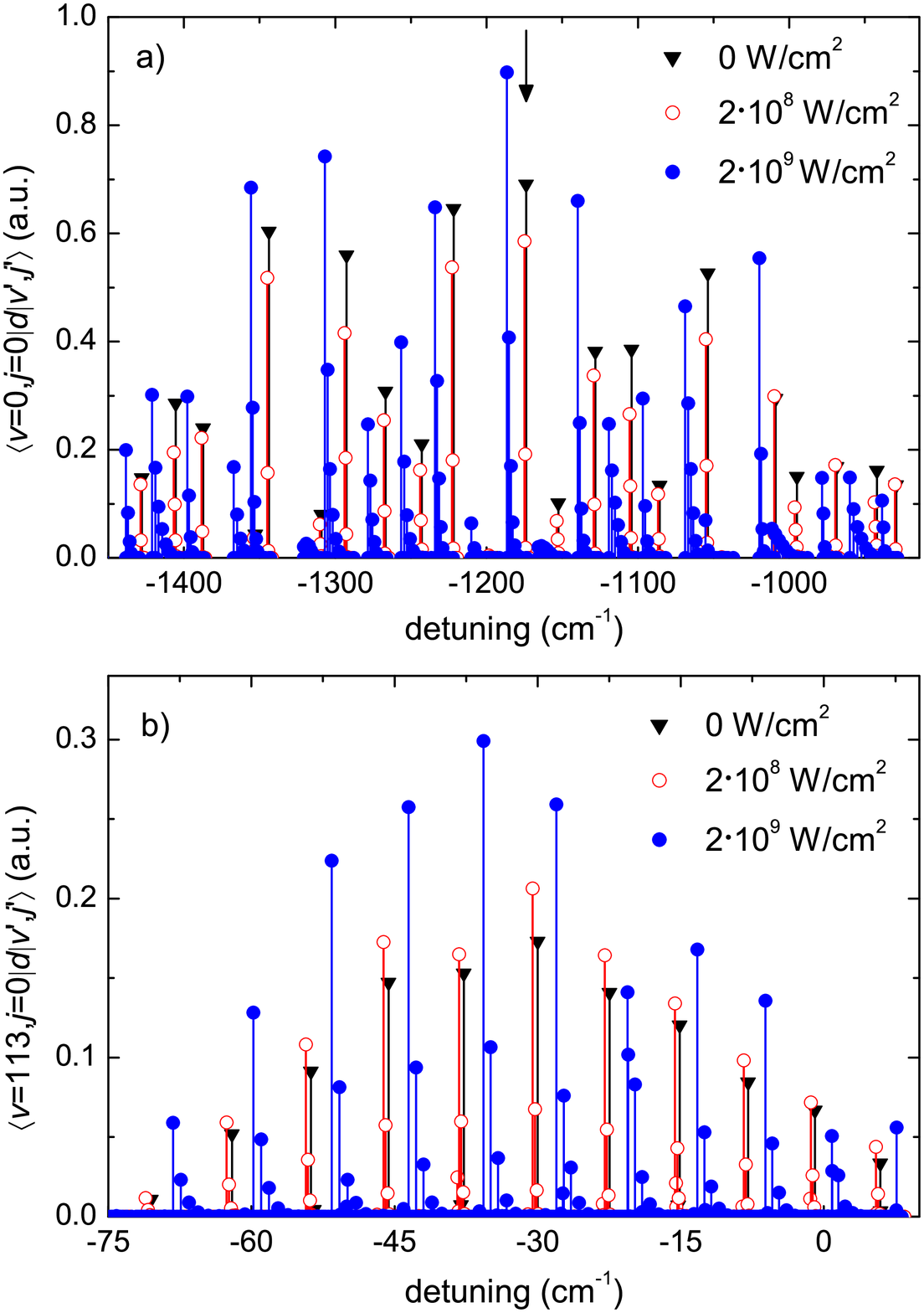}  
  \caption{
    Transition dipole matrix elements 
    between the ground rovibrational level $v=0,j=0$ $(a)$ and the highly
    excited level $v=113,j=0$ $(b)$ of the X$^1\Sigma_g^+$  ground electronic state and rovibrational
    levels of the A$^1\Sigma_u^+$ and b$^3\Pi_u$ manifold
    for three intensities of the non-resonant field in ${}^{87}$Rb$_2$. The binding
    energy of the field-free X$^1\Sigma_g^+$ $v=113,j=0$ level is
    $E_b=8.3\,$cm$^{-1}$. 
    The detuning is computed as 
    $E_{v',J'} - E_{v,J}-(E_{^2P_{1/2}}-E_{^2S})$, with $E_{^2P_{1/2}}$,
    $E_{^2S}$ the field-free energies of the atomic levels. 
  }
  \label{fig11}
\end{figure}
\begin{figure}[tbp]
  \centering
  \includegraphics[width=0.75\columnwidth]{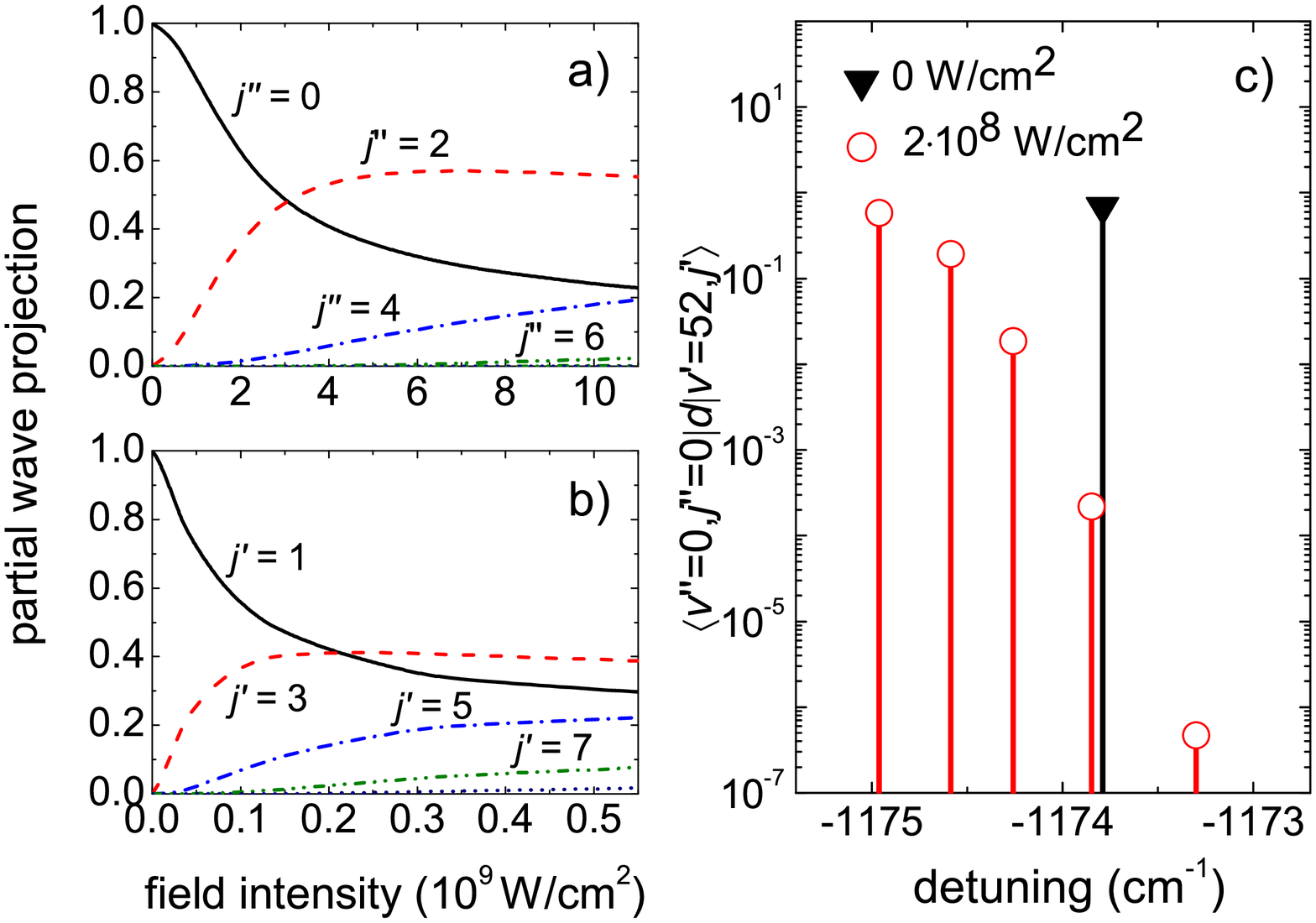}  
  \caption{Partial wave decomposition of the field-dressed
    rovibrational wavefunctions  for the X$^1\Sigma_g^+$ state
    $v=0,j=0$ ground level $(a)$ and the  
    $v'=52,J'=1$ level $(b)$ of the A$^1\Sigma_u^+$ and
    b$^3\Pi_u$ manifold in ${}^{87}$Rb$_2$. Also shown are the electric dipole
    transition moments between the X$^1\Sigma_g^+$ state
    $v=0,j=0$ ground level and the rotational manifold
    with $v'=52$ $(c)$.} 
  \label{fig12}
\end{figure}
Comparing three different intensities,
Fig.~\ref{fig11} illustrates the effect of the non-resonant field on
the transition dipole matrix elements for transitions between 
the X$^1\Sigma_g^+$ ground state and the A$^1\Sigma_u^+$ and
b$^3\Pi_u$ excited states. 
The transition dipole matrix elements are
calculated as rovibrational average of Eq.~(\ref{trandipel}) for 
given field-dressed rovibrational levels, i.e.,  
$\sum_{k=\mathrm{A}^1\Sigma_u^+,\mathrm{b}^3\Pi_u}
\left\langle \varphi^k_{v',j'}\big|
d_z(k\leftarrow \mathrm{X})(R)\cos\theta\big|
\varphi^{\mathrm{X}^1\Sigma_g^+}_{v,j}\right\rangle$, and shown
for the X$^1\Sigma_g^+$ state ground level in Fig.~\ref{fig11}$(a)$ 
and a vibrationally highly excited level in Fig.~\ref{fig11}$(b)$. 
These levels could be studied using molecules in a molecular beam $(a)$
or produced by photoassociation $(b)$~\cite{HyewonPRA07}.
Inspection of Fig.~\ref{fig11} reveals that the transitions get
shifted as expected, due to the decrease of all eigenenergies in the
non-resonant field~\cite{AganogluKoch,GonzalezPRA12}. Moreoever, the
transition strengths are strongly modified. This modification is
analyzed in more detail in Fig.~\ref{fig12} for the strongest
transition from the X$^1\Sigma_g^+$ state ground level 
indicated by an arrow in Fig.~\ref{fig11}$(a)$. Due to hybridization of
the rotational motion, illustrated in Fig.~\ref{fig12}$(a)$ and
$(b)$ in terms of the projections of the rovibrational wavefunctions
onto the field-free partial waves, the wavefunctions consist of
contributions from several field-free partial waves between which
transitions are allowed. This yields a series of rovibrational lines
observed in Fig.~\ref{fig12}$(c)$ instead of the single line for
$v=0,j=0$ to $v'=52,j'=1$ in the field-free case. 
For the largest intensity shown in Fig.~\ref{fig11}, $\mathcal{I}=2\cdot
10^{9}\,$W/cm$^2$, the transition matrix elements for the strongest
lines are clearly larger than in the field-free case. This is
rationalized by an alignment of the field-dressed
levels  in the ground and excited electronic states, with 
$\langle \cos^2\theta\rangle\gtrsim 0.73$ for $\mathcal{I}=2\cdot
10^{9}\,$W/cm$^2$. Correspondingly, 
the field-dressed  wavefunctions are localized in the angular
regions $\theta$  close to $0$ and $\pi$. As a consequence, the
field-dressed transition strengths are larger than the field-free ones due to the
angular dependence of the matrix
elements on $\cos\theta$~\cite{MaylePRA07}.

\begin{figure}[tbp]
  \centering
  \includegraphics[width=0.75\columnwidth]{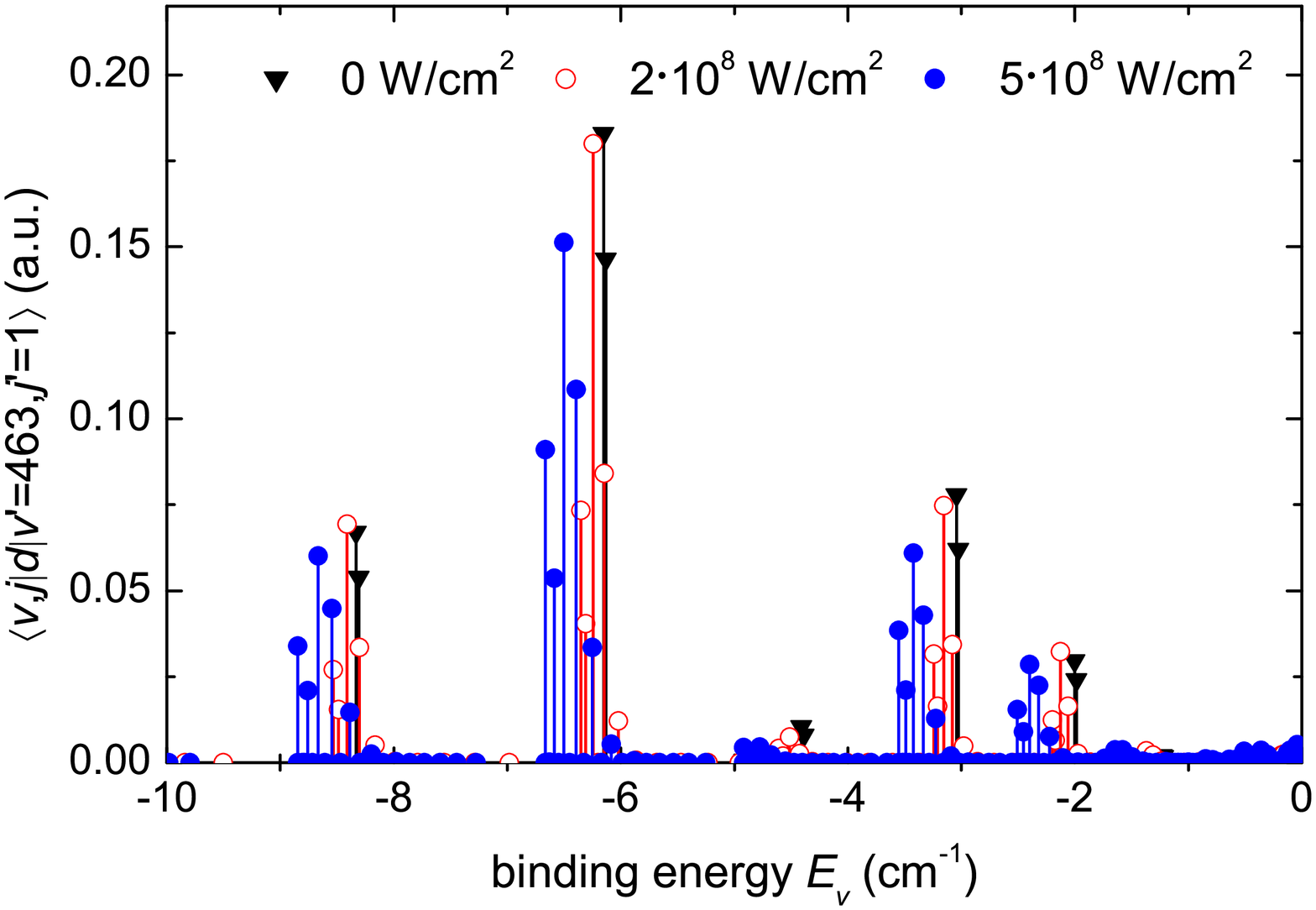}  
  \caption{
    Transition dipole matrix elements for a highly
    excited rovibrational level ($v'=463$,
    $E^{v'=463}_{bind}=8.3\,$cm$^{-1}$) of the A$^1\Sigma_u^+$ and b$^3\Pi_u$
    manifold and highly excited X$^1\Sigma_g^+$ state levels in ${}^{87}$Rb$_2$. 
  }
  \label{fig13}
\end{figure}
Figure~\ref{fig13} illustrates the effect of a non-resonant field on
the transition dipole moments for a weakly bound level in the excited 
A$^1\Sigma_u^+$ and b$^3\Pi_u$ state manifold. This level is
particularly well-suited for the photoassociative production of Rb$_2$
molecules~\cite{HyewonPRA07}, and the analysis of Fig.~\ref{fig13} is
motivated by a recent proposal for enhancing photoassociation rates
using a non-resonant field~\cite{GonzalezPRA12}. While the
calculations of Ref.~\cite{GonzalezPRA12} were carried out for Sr$_2$,
a somewhat smaller, albeit still significant enhancement of the
photoassociation rate of about one order of magnitude can be expected
for Rb$_2$~\cite{AganogluKoch}. The non-resonant field will affect the
spontaneous decay of the photoassociated molecules which is governed
by the matrix elements shown in Fig.~\ref{fig13}. The field-free data
represents a rotationally resolved equivalent of Fig.~3 of
Ref.~\cite{HyewonPRA07}. The binding energy of $8.3\,$cm$^{-1}$ in
Fig.~\ref{fig13} corresponds to the ground state level $v=113$,
cf. Fig.~\ref{fig11}$(b)$. A weak non-resonant field splits the two
lines originating from the $j'=1$ level into several ones, similar to
Fig.~\ref{fig12}$(c)$. The transition strength for $j=0$ is
almost not affected by the weak field. This behavior is similar to what
has been observed for transitions between weakly bound levels of the
strontium dimer~\cite{GonzalezPRA12}. For a strong non-resonant field,
the binding energies are shifted 
and the overall behavior is similar to
Ref.~\cite{GonzalezPRA12}. This implies that a non-resonant field may
enhance the photoassociation rate without compromising an efficient 
stabilization into bound ground state levels by spontaneous emission
as it was observed in Ref.~\cite{HyewonPRA07}.

\begin{figure}[tbp]
  \centering
  \includegraphics[width=0.75\columnwidth]{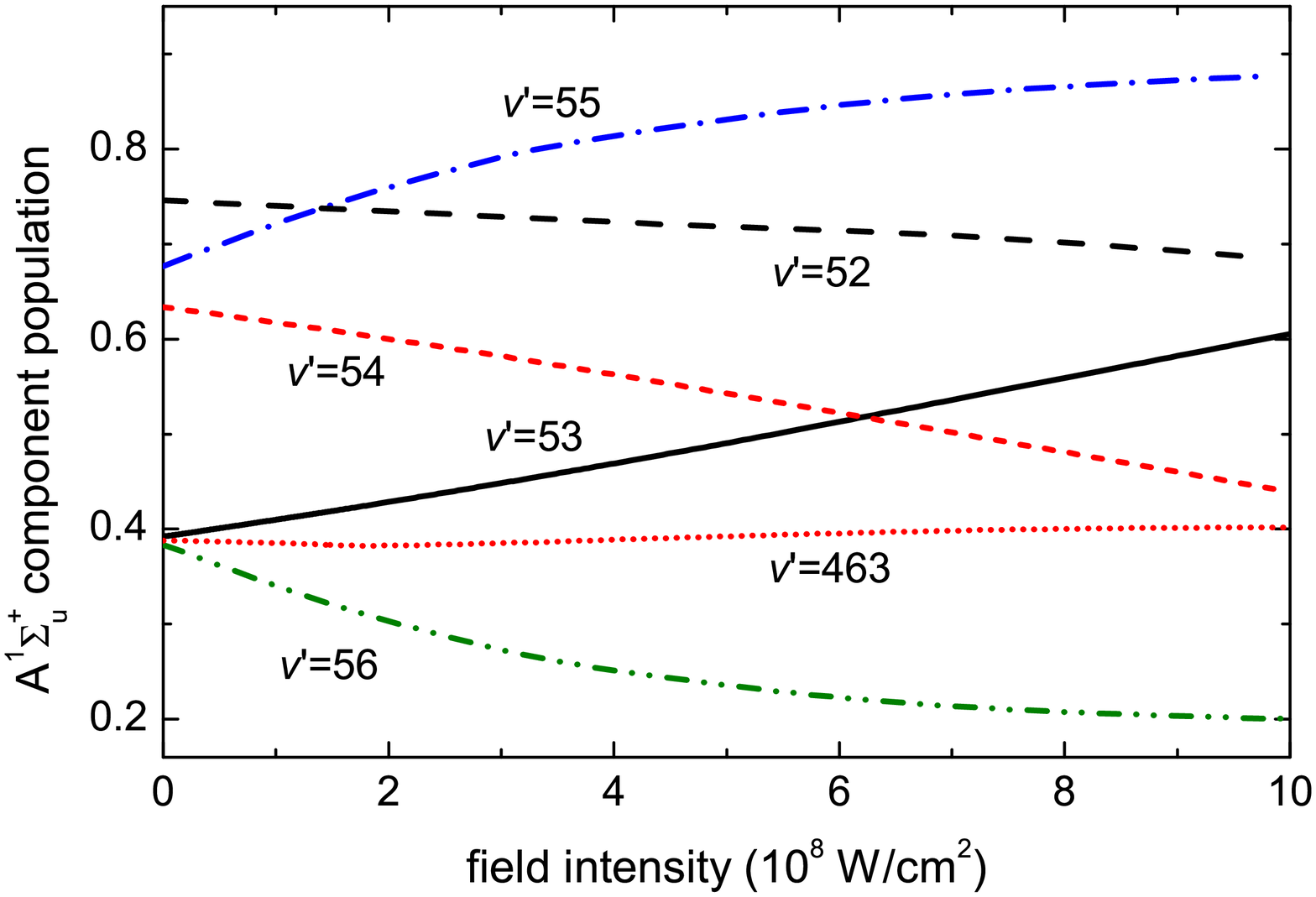}  
  \caption{Singlet component of the coupled excited state 
    rovibrational levels vs non-resonant field intensity with $v'$
    the field-free vibrational quantum number. Data shown for
    $j=1$ (the behavior for other $j$ is very similar).} 
  \label{fig14}
\end{figure}
Finally, Fig.~\ref{fig14} analyzes the interplay of the spin-orbit
coupling and the interaction with the non-resonant field for several 
of the rovibrational levels of the A$^1\Sigma_u^+$ and b$^3\Pi_u$
manifold studied in Figs.~\ref{fig11}$(a)$, \ref{fig12} and \ref{fig13}. 
Surprisingly, the levels from the middle of the well, $v'=52,\ldots,56$,
show a remarkable dependence of the singlet-triplet decomposition on
the non-resonant field intensity. On the other hand, the
singlet-triplet character of weakly bound
levels of the A$^1\Sigma_u^+$ and b$^3\Pi_u$ manifold, shown here for
the representative $v'=463$, is hardly affected. 
This behavior can be understood by inspection of the $R$-dependence of
the polarizability components and the spin-orbit coupling matrix
elements, cf. Figs.~\ref{fig9} and \ref{fig7}.
Weakly bound levels have most of their amplitude at internuclear
separations larger than $R=20\,$a$_0$. The spin-orbit coupling is
strong at large internuclear separations and smaller at intermediate
separations, while the opposite is true for the polarizability
components. A large dependence of the singlet-triplet character on the
non-resonant field intensity is expected when the interaction energy
with the field and the spin-orbit coupling become comparable. Due to
the $R$-dependence of the polarizability, for weakly bound levels this
requires field 
intensities in excess of  $10^{10}\,$W/cm$^2$. On the other hand, the
more deeply bound levels, $v'=52,\ldots,56$, have their outer
turning point near $R=12\,a_0$ where the polarizability is large and the
spin-orbit coupling is small. Therefore, intensities of the order of
$10^{9}\,$W/cm$^2$ yield an interaction energy with the field that is
comparable to  the spin-orbit coupling. For example, for
$10^9\,$W/cm$^2$, the Stark shift of the levels amounts to about 
15$\,$cm$^{-1}$. Their vibrational spacing, of the order of
20$\,$cm$^{-1}$,  is also comparable. The interaction with the
non-resonant field will then affect the singlet-triplet character of a
rovibrational wavefunction, provided the $R$-dependence of
polarizabilities differs for singlet and triplet states. This is
indeed the case, cf. Fig.~\ref{fig9}, explaining the changes in the
singlet-triplet decomposition observed in Fig.~\ref{fig14}.

\section{Summary and conclusions}
\label{sec:concl}

In the present work we have investigated how the spectroscopy of the
Rb$_2$ molecule is affected by applying a non-resonant field. Our
emphasis has been on the manifold of  the spin-orbit coupled
A$^1\Sigma_u^+$ and b$^3\Pi_u$ excited states. To this end we have
derived the electronic Hamiltonian describing the interaction of a
diatomic molecule with a non-resonant field in general and 
the Hamiltonian describing the nuclear
motion in a non-resonant field for the manifold of the coupled
A$^1\Sigma_u^+$ and b$^3\Pi_u$ excited electronic states in
particular. We have employed the double
electron attachment intermediate Hamiltonian Fock space coupled
cluster method restricted to single and double excitations
for all electronic states of the Rb$_2$ molecule up to the $5s+5d$
dissociation limit of about 26.000$\,$cm$^{-1}$. The agreement between
the present results 
and those fitted to high resolution spectroscopic data is very good, both
for the well depths and the vibrational frequencies.
The accuracy of the present results for the potential energy
curves is much higher than the previous electronic structure calculations in Refs.~\cite{Park:01,Edwardsson:03}
and slightly better than in the most recent study by Allouche and
Aubert-Fr\'econ \cite{Allouche:12}.

In order to correctly predict the spectroscopic behavior, we
have also calculated the electric transition dipole moments,
non-adiabatic coupling and spin-orbit coupling matrix elements, and
static dipole polarizabilities of Rb$_2$, using the multireference
configuration interaction method. To the best of our knowledge, we
have reported in this paper the very first calculation of the
irreducible components of the polarizability tensor as 
a function of $R$ for electronically excited states.
For the spin-orbit coupled manifold of the A$^1\Sigma_u^+$ and b$^3\Pi_u$ 
excited states, we have checked the accuracy of the {\em ab initio}
results with the spectroscopic data. Very good agreement was found.

We have investigated the spectroscopy of Rb$_2$ in its rovibronic
ground state, corresponding to a molecular beam experiment, as well as
in highly excited vibrational levels of ground and electronically
excited states, typical for photoassociation
experiments at ultracold temperatures. In both cases, 
the spectroscopy is significantly altered by a non-resonant
field. Specifically, fields of the order of $10^8\,$W/cm$^2$ are found
to split a single rovibrational line into several ones and shift the lines 
by a few cm$^{-1}$. The splitting is due to rotational hybridization,
i.e., the field-dressed wavefunctions are made up of several 
field-free partial waves with comparable contributions. For strong
non-resonant fields, of the order of $10^9\,$W/cm$^2$, alignment leads
to an increase of the transition strengths compared to the field-free
case, due to localization of the rotational wavefunctions in regions
close to $\theta=0$ and $\pi$, and the dependence of the transition
matrix elements on $\cos\theta$~\cite{MaylePRA07}. 
We have also investigated the effect of a non-resonant field on the
transition matrix elements that govern stabilization by spontaneous
emission for photoassociated molecules~\cite{HyewonPRA07}. Similarly
to strontium molecules~\cite{GonzalezPRA12}, transitions occur to the
same vibrational levels as in the field-free case. This implies that a
non-resonant field may be used to enhance the photoassociation
rate~\cite{GonzalezPRA12} without deteriorating stabilization of the
photoassociated molecules into  bound levels of the electronic ground
state. Somewhat surprisingly, we have found
a non-resonant field to significantly modify the singlet-triplet
character of rovibrational levels in the A$^1\Sigma_u^+$ and
b$^3\Pi_u$ excited state manifold for levels in the middle of the
potential wells, while weakly bound levels remain rather
unaffected. We have identified two conditions for a modification of
the singlet-triplet character -- the interaction energy with the field
needs to be comparable to the spin-orbit coupling and the dependence
of the polarizability tensor components on the interatomic separation
must differ for singlet and triplet molecules. If both conditions are
fulfilled, as was found to be the case for Rb$_2$ levels of the
A$^1\Sigma_u^+$ and b$^3\Pi_u$ manifold  with vibrational
quantum number around 55, the singlet or triplet character of a
rovibrational wavefunction can be controlled by a non-resonant field. 

An interesting perspective for coherent control arises when applying
a non-resonant field to degenerate excited electronic states. We have
shown that, for degenerate states, a non-resonant field introduces a
coupling between different states, $0_u^+$ and $2_u$ in the present
example. In coherent control based on wavepacket motion, such 
a coupling between different states can be used to shape the
effective potential energy curve governing the wavepacket
dynamics~\cite{KochPRA08}. Using a non-resonant field, for example in
the far infrared, comes with the advantage of small losses even for
strong fields. Non-resonant field control of photoassociation
rates~\cite{GonzalezPRA12} or wavepacket dynamics based on
field-induced resonant coupling~\cite{KochPRA08}
 represents a new twist to manipulating molecules
with non-resonant fields~\cite{FriedrichPRL95}.

\section*{Acknowledgements}

Financial support from the Polish Ministry of Science and
Higher Education through the project N N204 215539
and   by the Spanish project FIS2011-24540 (MICINN) as well as the
Grants P11-FQM-7276 and FQM-4643 (Junta de Andaluc\'{\i}a) is
gratefully acknowledged. 
MT was supported by the project operated within the Foundation for
Polish Science MPD Programme co-financed by the EU European Regional
Development Fund. RGF belongs to the Andalusian research group
FQM-207. RM thanks the Foundation for Polish Science for support
within the MISTRZ programme. 
Part of this work was done while the authors were visitors at 
the Kavli Institute for Theoretical Physics, University of California 
at Santa Barbara within the programme Fundamental Science and Applications of
Ultra-cold Polar Molecules. Financial support from the National Science 
Foundation grant no. NSF PHY11-25915 is gratefully acknowledged.


\begin{thebibliography}{79}
\providecommand{\url}[1]{\texttt{#1}}
\providecommand{\urlprefix}{URL }
\markboth{Taylor \& Francis and I.T. Consultant}{Molecular Physics}

\bibitem{Cornell96}
E. Cornell,  J. Res. Natl. Inst. Stand. Techol.  \textbf{101}, 419 (1996).

\bibitem{RobertsPRL98}
J.L. Roberts, N. Claussen, J.P. Burke~Jr., C.H. Greene, E.A. Cornell and C.E.
  Wieman,  Phys. Rev. Lett.  \textbf{81} (23), 5109 (1998).

\bibitem{vanKempenPRL02}
E.G.M. van Kempen, S.J.J.M.F. Kokkelmans, D.J. Heinzen and B.J. Verhaar,  Phys.
  Rev. Lett.  \textbf{88}, 093201 (2002).

\bibitem{MartePRL02}
A. Marte, T. Volz, J. Schuster, S. D\"urr, G. Rempe, E.G.M. van Kempen and B.J.
  Verhaar,  Phys. Rev. Lett.  \textbf{89}, 283202 (2002).

\bibitem{GabbaniniPRL00}
C. Gabbanini, A. Fioretti, A. Lucchesini, S. Gozzini and M. Mazzoni,  Phys.
  Rev. Lett.  \textbf{84} (13), 2814 (2000).

\bibitem{DuerrPRL04}
S. D\"urr, T. Volz, A. Marte and G. Rempe,  Phys. Rev. Lett.  \textbf{92},
  020406 (2004).

\bibitem{BoestenPRL96}
H.M.J.M. Boesten, C.C. Tsai, B.J. Verhaar and D.J. Heinzen,  Phys. Rev. Lett.
  \textbf{77}, 5194 (1996).

\bibitem{BoestenPRA97}
H.M.J.M. Boesten, C.C. Tsai, J.R. Gardner, D.J. Heinzen and B.J. Verhaar,
  Phys. Rev. A  \textbf{55} (1), 636 (1997).

\bibitem{VolzPRA05}
T. Volz, S. D\"urr, N. Syassen, G. Rempe, E. van Kempen and S. Kokkelmans,
  Phys. Rev. A  \textbf{72}, 010704 (2005).

\bibitem{WinklerPRL05}
K. Winkler, G. Thalhammer, M. Theis, H. Ritsch, R. Grimm and J.H. Denschlag,
  Phys. Rev. Lett.  \textbf{95}, 063202 (2005).

\bibitem{LangPRL08}
F. Lang, K. Winkler, C. Strauss, R. Grimm and J.H. Denschlag,  Phys. Rev. Lett.
   \textbf{101}, 133005 (2008).

\bibitem{WrightPRL05}
M.J. Wright, S.D. Gensemer, J. Vala, R. Kosloff and P.L. Gould,  Phys. Rev.
  Lett.  \textbf{95}, 063001 (2005).

\bibitem{WrightPRA06}
M.J. Wright, J.A. Pechkis, J.L. Carini and P.L. Gould,  Phys. Rev. A
  \textbf{74}, 063402 (2006).

\bibitem{WrightPRA07}
M.J. Wright, J.A. Pechkis, J.L. Carini, S. Kallush, R. Kosloff and P.L. Gould,
  Phys. Rev. A  \textbf{75}, 051401 (2007).

\bibitem{PechkisPRA11}
J.A. Pechkis, J.L. Carini, C.E. Rogers, P.L. Gould, S. Kallush and R. Kosloff,
  Phys. Rev. A  \textbf{83} (6), 063403 (2011).

\bibitem{CariniPRA13}
J.L. Carini, J.A. Pechkis, C.E. Rogers, P.L. Gould, S. Kallush and R. Kosloff,
  Phys. Rev. A  \textbf{87}, 011401 (2013).

\bibitem{SalzmannPRL08}
W. Salzmann, T. Mullins, J. Eng, M. Albert, R. Wester, M. Weidem\"{u}ller, A.
  Merli, S.M. Weber, F. Sauer, M. Plewicki, F. Weise, L. W\"{o}ste and A.
  Lindinger,  Phys. Rev. Lett.  \textbf{100}, 233003 (2008).

\bibitem{MullinsPRA09}
T. Mullins, W. Salzmann, S. G\"otz, M. Albert, J. Eng, R. Wester, M.
  Weidem\"uller, F. Weise, A. Merli, S.M. Weber, F. Sauer, L. W\"oste and A.
  Lindinger,  Phys. Rev. A  \textbf{80} (6), 063416 (2009).

\bibitem{MerliPRA09}
A. Merli, F. Eimer, F. Weise, A. Lindinger, W. Salzmann, T. Mullins, S.
  G\"{o}tz, R. Wester, M. Weidem\"{u}ller, R. A\u{g}ano\u{g}lu and C.P. Koch,
  Phys. Rev. A  \textbf{80}, 063417 (2009).

\bibitem{McCabePRA09}
D.J. McCabe, D.G. England, H.E.L. Martay, M.E. Friedman, J. Petrovic, E.
  Dimova, B. Chatel and I.A. Walmsley,  Phys. Rev. A  \textbf{80}, 033404
  (2009).

\bibitem{BergemanJPhysB06}
T. Bergeman, J. Qi, D. Wang, Y. Huang, H.K. Pechkis, E.E. Eyler, P.L. Gould,
  W.C. Stwalley, R.A. Cline, J.D. Miller and D.J. Heinzen,  J. Phys. B
  \textbf{39}, S813 (2006).

\bibitem{HuangJPhysB06}
Y. Huang, J. Qi, H.K. Pechkis, D. Wang, E.E. Eyler, P.L. Gould and W.C.
  Stwalley,  J. Phys. B  \textbf{39}, S857 (2006).

\bibitem{HyewonPRA07}
H.K. Pechkis, D. Wang, Y. Huang, E.E. Eyler, P.L. Gould, W.C. Stwalley and C.P.
  Koch,  Phys. Rev. A  \textbf{76}, 022504 (2007).

\bibitem{FiorettiJPB07}
A. Fioretti, O. Dulieu and C. Gabbanini,  J. Phys. B  \textbf{40} (16), 3283
  (2007).

\bibitem{BellosPCCP11}
M.A. Bellos, D. Rahmlow, R. Carollo, J. Banerjee, O. Dulieu, A. Gerdes, E.E.
  Eyler, P.L. Gould and W.C. Stwalley,  Phys. Chem. Chem. Phys.  \textbf{13},
  18880 (2011).

\bibitem{Huber:79}
P.H. Huber and G. Herzberg, \emph{Molecular Spectra and Molecular Structure.
  IV. Constants of Diatomic Molecules}   (Van Nostrand Reinhold Company, New
  York, 1979).

\bibitem{LawrencePR29}
E.O. Lawrence and N.E. Edlefsen,  Phys. Rev.  \textbf{34}, 233 (1929).

\bibitem{LeRoy:00}
J.Y. Seto, R.J.L. Roy, J. Verg\`{e}s and C. Amiot,  J. Chem. Phys.
  \textbf{113} (8), 3067 (2000).

\bibitem{Lozeille:06}
J. Lozeille, A. Fioretti, C. Gabbanini, Y. Huang, H.K. Pechkis, D. Wang, P.L.
  Gould, E.E. Eyler, W.C. Stwalley, M. Aymar and O. Dulieu,  Eur. Phys. J. D
  \textbf{39}, 261 (2006).

\bibitem{Beser:09}
B. Beser, V.B. Sovkov, J. Bai, E.H. Ahmed, C.C. Tsai, F. Xie, L. Li, V.S.
  Ivanov and A.M. Lyyra,  J. Chem. Phys.  \textbf{131}, 094505 (2009).

\bibitem{Tiemann:10}
C. Strauss, T. Takekoshi, F. Lang, K. Winkler, R. Grimm, J. Hecker~Denschlag
  and E. Tiemann,  Phys. Rev. A  \textbf{82}, 052514 (2010).

\bibitem{Mudrich:09}
M. Mudrich, P. Heister, T. Hippler, C. Giese, O. Dulieu and F. Stienkemeier,
  Phys. Rev. A  \textbf{80}, 042512 (2009).

\bibitem{Aubock:10}
G. Aub\"{o}ck, M. Aymar, O. Dulieu and W.E. Ernst,  J. Chem. Phys.
  \textbf{132}, 054304 (2010).

\bibitem{GutterresPRA02}
R.F. Gutterres, C. Amiot, A. Fioretti, G. Gabbanini, M. Mazzoni and O. Dulieu,
  Phys. Rev. A  \textbf{66}, 024502 (2002).

\bibitem{SalamiPRA09}
H. Salami, T. Bergeman, B. Beser, J. Bai, E.H. Ahmed, S. Kotochigova, A.M.
  Lyyra, J. Huennekens, C. Lisdat, A.V. Stolyarov, O. Dulieu, P. Crozet and
  A.J. Ross,  Phys. Rev. A  \textbf{80}, 022515 (2009).

\bibitem{Amiot:86}
C. Amiot,  Mol. Phys.  \textbf{58} (4), 667 (1986).

\bibitem{Amiot:87}
C. Amiot and J. Verges,  Mol. Phys.  \textbf{61} (1), 51 (1987).

\bibitem{Amiot:90}
C. Amiot,  J. Chem. Phys.  \textbf{93} (12), 8591 (1990).

\bibitem{KonowalowACS82}
D. Konowalow and M. Rosenkrantz,  {ACS Symp. Ser.}  \textbf{{179}}, {3}
  ({1982}).

\bibitem{Park:01}
S.J. Park, S.W. Suh, Y.S. Lee and G.H. Jeung,  J. Mol. Spectrosc.  \textbf{207}
  (2), 129 (2001).

\bibitem{Edwardsson:03}
D. Edvardsson, S. Lunell and C.M. Marian,  Mol. Phys.  \textbf{101}, 2381
  (2003).

\bibitem{Allouche:12}
A.R. Allouche and M. Aubert-Frecon,  J. Chem. Phys.  \textbf{136} (11), 114302
  (2012).

\bibitem{Jastrzebski}
W. Jastrzebski private communication.

\bibitem{Gould}
P.L. Gould private communication.

\bibitem{BanCPL01}
T. Ban, H. Skenderovi\'c, R. Beuc, I.K. Broni\'c, S. Rousseau, A. Allouche, M.
  Aubert-Fr\'econ and G. Pichler,  Chem. Phys. Lett.  \textbf{345}, 423 (2001).

\bibitem{BanEPL04}
T. Ban, R. Beuc, H. Skenderović and G. Pichler,  Europhys. Lett.  \textbf{66}
  (4), 485 (2004).

\bibitem{TomzaPRA12}
M. Tomza, M.H. Goerz, M. Musia\l{}, R. Moszynski and C.P. Koch,  Phys. Rev. A
  \textbf{86}, 043424 (2012).

\bibitem{KochFaraday09}
C.P. Koch, M. Ndong and R. Kosloff,  Faraday Disc.  \textbf{142}, 389 (2009).

\bibitem{MusialJCP12}
M. Musia\l,  The Journal of Chemical Physics  \textbf{136} (13), 134111 (2012).

\bibitem{MusialRMP07}
R.J. Bartlett and M. Musia\l{},  Rev. Mod. Phys.  \textbf{79}, 291 (2007).

\bibitem{MusialCR12}
D.I. Lyakh, M. Musia\l{}, V.F. Lotrich and R.J. Bartlett,  Chemical Reviews
  \textbf{112} (1), 182 (2012).

\bibitem{SpelsbergJCP99}
D. Spelsberg,  {J. Chem. Phys.}  \textbf{{111}}, {9625} ({1999}).

\bibitem{SkomorowskiJCP11}
W. Skomorowski and R. Moszynski,  J. Chem. Phys.  \textbf{134} (12), 124117
  (2011).

\bibitem{AganogluKoch}
R. A\u{g}ano\u{g}lu, M. Lemeshko, B. Friedrich, R. Gonz\'alez-F\'erez and C.P.
  Koch,  arXiv:1105.0761   (2011).

\bibitem{GonzalezPRA12}
R. Gonz\'alez-F\'erez and C.P. Koch,  Phys. Rev. A  \textbf{86}, 063420 (2012).

\bibitem{HeijmenMP96}
T.G.A. Heijmen, R. Moszynski, P.E.S. Wormer and A. van~der Avoird,  Mol. Phys.
  \textbf{89}, 81 (1996).

\bibitem{BusseryPRA03}
B. Bussery-Honvault, J.M. Launay and R. Moszynski,  Phys. Rev. A  \textbf{68},
  032718 (2003).

\bibitem{Moszynski:05}
B. Bussery-Honvault, J.M. Launay and R. Moszynski,  Phys. Rev. A  \textbf{72},
  012702 (2005).

\bibitem{BusseryJCP06}
B. Bussery-Honvault, J.M. Launay, T. Korona and R. Moszynski,  J. Chem. Phys.
  \textbf{125}, 114315 (2006).

\bibitem{BusseryMolPhys06}
B. Bussery-Honvault and R. Moszynski,  Mol. Phys.  \textbf{104} (13-14), 2387
  (2006).

\bibitem{KochPRA08}
C.P. Koch and R. Moszy\'nski,  Phys. Rev. A  \textbf{78}, 043417 (2008).

\bibitem{RybakPRL11}
L. Rybak, S. Amaran, L. Levin, M. Tomza, R. Moszynski, R. Kosloff, C.P. Koch
  and Z. Amitay,  Phys. Rev. Lett.  \textbf{107}, 273001 (2011).

\bibitem{RybakFaraday11}
L. Rybak, Z. Amitay, S. Amaran, R. Kosloff, M. Tomza, R. Moszynski and C.P.
  Koch,  Faraday Disc.  \textbf{153}, 383 (2011).

\bibitem{SkomorowskiPRA12}
W. Skomorowski, R. Moszynski and C.P. Koch,  Phys. Rev. A  \textbf{85}, 043414
  (2012).

\bibitem{SkomorowskiJCP12}
W. Skomorowski, F. Paw{\l}owski, C.P. Koch and R. Moszynski,  J. Chem. Phys.
  \textbf{136}, 194306 (2012).

\bibitem{KrychPRA11}
M. Krych, W. Skomorowski, F. Paw\l{}owski, R. Moszynski and Z. Idziaszek,
  Phys. Rev. A  \textbf{83}, 032723 (2011).

\bibitem{TomzaPCCP11}
M. Tomza, F. Paw{\l}owski, M. Jeziorska, C.P. Koch and R. Moszynski,  Phys.
  Chem. Chem. Phys.  \textbf{13} (42), 18893 (2011).

\bibitem{Boys:70}
S. Boys and F. Bernardi,  Mol. Phys.  \textbf{19}, 553 (1970).

\bibitem{MarinescuPRA95}
M. Marinescu and A. Dalgarno,  Phys. Rev. A  \textbf{52} (1), 311 (1995).

\bibitem{Bunker98}
P. Bunker and P. Jensen, \emph{Molecular Symmetry and Spectroscopy}   (NRC
  Press, Ottawa, 1998).

\bibitem{LimJCP05}
I.S. Lim, P. Schwerdtfeger, B. Metz and H. Stoll,  J. Chem. Phys.  \textbf{122}
  (10), 104103 (2005).

\bibitem{NIST}
{NIST Atomic Spectra Database,} http://physics.nist.gov/PhysRefData/ASD.

\bibitem{ACESII}
J.F. Stanton, J. Gauss, S.A. Perera, J.D. Watts, A. Yau, M. Nooijen, N.
  Oliphant, P. Szalay, W. Lauderdale, S. Gwaltney, S. Beck, A. Balkov, D.
  Bernholdt, K. Baeck, P. Rozyczko, H. Sekino, C. Huber, J. Pittner, W. Cencek,
  D. Taylor,  and R. Bartlett, {ACES II is a program product of the Quantum
  Theory Project, University of Florida} .

\bibitem{Molpro}
H.J. Werner, P.J. Knowles, F.R.M. R.~Lindh, M. Sch{\"u}tz, P. Celani, T.
  Korona, A. Mitrushenkov, G. Rauhut, T.B. Adler, R.D. Amos, A. Bernhardsson,
  A. Berning, D.L. Cooper, M.J.O. Deegan, A.J. Dobbyn, E.G. F.~Eckert, C.
  Hampel, G. Hetzer, T. Hrenar, G. Knizia, C. K{\"o}ppl, Y. Liu, A.W. Lloyd,
  R.A. Mata, A.J. May, S.J. McNicholas, W. Meyer, M.E. Mura, A. Nicklass, P.
  Palmieri, K. Pfl{\"u}ger, R. Pitzer, M. Reiher, U. Schumann, H. Stoll, A.J.
  Stone, R. Tarroni, T. Thorsteinsson, M. Wang and A. Wolf, \emph{MOLPRO,
  version 2008.1, a package of ab initio programs} 2008, See
  http://www.molpro.net.

\bibitem{FreyJPB78}
P. Frey, F. Breyer and H. Holop,  J. Phys. B: At. Mol. Phys.  \textbf{11} (19),
  L589 (1978).

\bibitem{Derevianko:10}
A. Derevianko, S.G. Porsev and J.F. Babb,  At. Data Nucl. Data Tables
  \textbf{96} (3), 323  (2010).

\bibitem{DeiglmayrJCP08}
J. Deiglmayr, M. Aymar, R. Wester, M. Weidemuller and O. Dulieu,  J. Chem.
  Phys.  \textbf{129} (6), 064309 (2008).

\bibitem{MaylePRA07}
M. Mayle, R. Gonz\'alez-F\'erez and P. Schmelcher,  Phys. Rev. A  \textbf{75},
  013421 (2007).

\bibitem{FriedrichPRL95}
B. Friedrich and D. Herschbach,  Phys. Rev. Lett.  \textbf{74}, 4623 (1995).

\end{thebibliography}
\end{document}